\documentclass[a4paper,12pt]{article}
\usepackage{epsfig}
\usepackage[american]{babel}
\usepackage{amssymb}

\begin{document}
\title{On possibility of realization of the Mandelbrot set
in coupled continuous systems}

\author{Olga~B.~Isaeva$^*$, Sergey~P.~Kuznetsov}
\date{}
\maketitle\begin{center} \emph{ Institute of Radio-Engineering and
Electronics of RAS, Saratov Branch, \\ Zelenaya 38, Saratov,
410019, Russia\\
$^*$E-mail:IsaevaOB@info.sgu.ru}\end{center}

\maketitle
\begin{abstract}
According to the method, suggested in our previous
work~(arxiv:nlin.CD/0509012) and based on the consideration of the
specially coupled systems, the possibility of physical realization
of the phenomena of complex analytic dynamics (such as Mandelbrot
and Julia sets) is discussed. It is shown, that unlike the case of
discrete maps or differential systems with periodic driving,
investigated in mentioned work, there are some difficulties in
attempts to obtain the Mandelbrot set for the coupled autonomous
continuous systems. A system of coupled autonomous R\"{o}ssler
oscillators is considered as an example.
\end{abstract}

\section{Introduction}
It is known~\cite{Peitgen,Devaney}, that complex analytic dynamics
(CAD), studying behavior of complex maps, includes a lot of
interesting phenomena, for example, presence of fractal Mandelbrot
and Julia sets in the parameter and phase spaces.

Let us start with a quadratic logistic map
\begin{equation}\label{eq1}
{z}' \to \lambda - z^{2},
\end{equation}
where $\lambda$ is a complex parameter, and $z$ is a complex
variable. By definition, Mandelbrot set (fig.~1$à$) is a set of
points on a plane of complex parameter $\lambda$, for which the
orbit of an extremum  $z=0$ of the map~(\ref{eq1}) during
iteration procedure does not escape to infinity. The Mandelbrot
set contains the so-called "Mandelbrot cactus" (designed on figure
by gray color); this is a set of points in the parameter plane,
for which the trajectory starting of the extremum of the map
converges to a periodic attractor.

\begin{figure}
\centerline{\epsfig{file=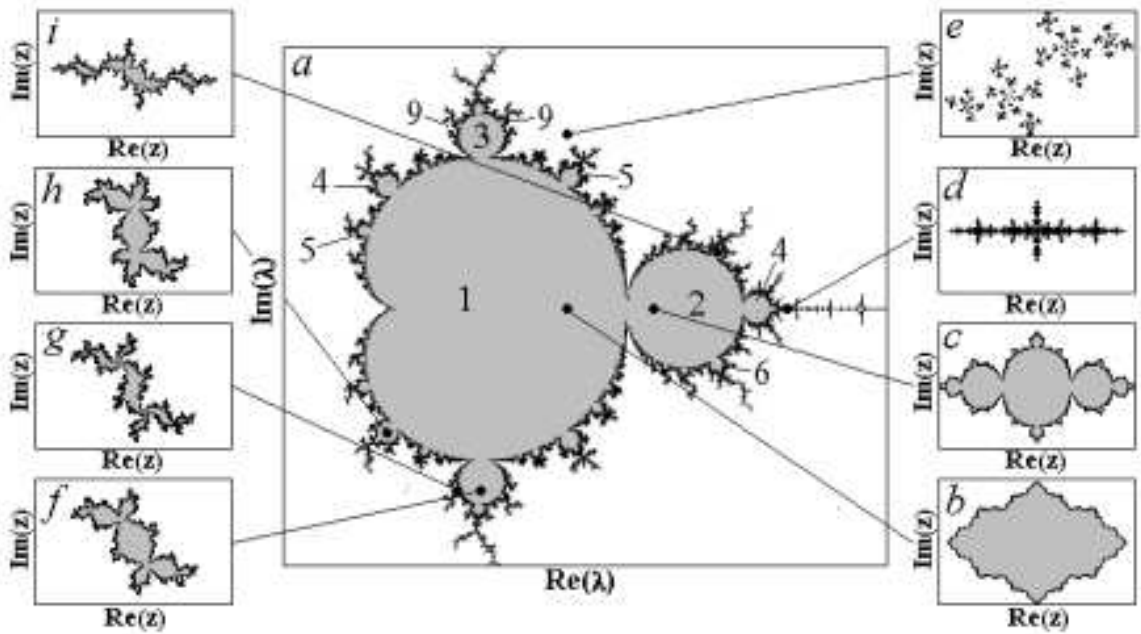,width=0.85\textwidth}}

\caption{Mandelbrot set (a) and Julia sets for the quadratic
complex map with different values of parameter: $\lambda=0.5$ (b),
$\lambda=0.8$ (c), $\lambda=1.42$ (d), $\lambda=0.5+0.7i$ (e),
$\lambda=0.123-0.745i$ (f), $\lambda=0.0315-0.7908i$ (g),
$\lambda=-0.282+0.530i$ (h), $\lambda=1.16+0.25i$ (i). The gray
color designates the regions corresponding to existence of
periodic dynamics (periods are marked by respective numbers); the
black color designates points, at which the restricted in a phase
space chaotic dynamics is implemented; the white color means the
escaping of trajectories to infinity.} \label{fig1}
\end{figure}

"Mandelbrot cactus" consists of a big cardioid, corresponding to
the existence of an attracting fixed point, and an infinite number
of "leaves", corresponding to existence of attracting cycles of
different periods. For example, the leaves of the doubled periods
are placed along a real axis. The sequences of other period
m-tupling (period-multiplication) bifurcations also can be found.
In particular, the accumulation points for period-tripling and
period-quadrupling bifurcation cascades have been investigated in
the work of Golberg, Sinai and Khanin~\cite{Golberg}. In the works
of Cvitanovi\'{c} and Mirheim~\cite{Cvitanovic1,Cvitanovic2} the
universal properties of many other bifurcation cascades were
studied.

A bifurcation, which is responsible for originating a "leave",
corresponds to a stability loss of a "parent" cycle, characterized
by a complex multiplier with unit modulus and rational argument
(in relation to $2\pi$). If the argument of a multiplier at the
stability loss is irrational, then the domains with fractal
boundaries, filled by invariant curves arise in the phase plane,
the so-called Siegel disks~\cite{Widom,Manton,MacKay}.

Fractal pattern close to the "Mandelbrot cactus" and denoted by
black color in Fig.~1, corresponds to existence of the chaotic
dynamical regimes in the phase space.

In figures~1($b$-$i$) the Julia sets for different values of
complex parameter $\lambda$ are shown. The Julia set is a border
between basins of attraction to infinity (white color) and to a
periodic motion (gray color) in a plane of complex variable $z$.
One can distinguish the following types of Julia sets:

\begin{itemize}
  \item for values $\lambda$, belonging to the "Mandelbrot cactus", the Julia
set is connected and enclose an interior basin
(figs.~1($b$,$c$,$f$-$i$);
  \item for values $\lambda$, at which chaotic dynamics exists, the Julia
set is also connected, but has no inner region (fig.~1$d$);
  \item for values $\lambda$, outside the Mandelbrot set, the Julia set is
disconnected (fig.~1$f$).
\end{itemize}

It is obvious, that 1D complex map can be represented equivalently
by a 2D real map (for this purpose it is necessary only to
separate real and imaginary parts of the equation). However, the
mentioned phenomena of CAD are intrinsic only to a very special
class of the real 2D maps, namely for the analytic maps, obeying
the Cauchy-Riemann conditions. Violation of the analyticity leads
to drastic changes of the dynamics of the
map~\cite{rcd,Peinke,Klein,Peckham1,Peckham2}. Thereby, a
following problem arises: Is it possible to specify actual
physical systems demonstrating phenomena of CAD? Recently, this
problem attracts great attention. The physical applications of
complex dynamics for such problems, as the renormalization group
approach in the theory of phase transitions and the theory of a
percolation were discussed~\cite{Hu,percol,Abdusalam,npcs}. In the
paper of Beck~\cite{Beck} a theoretical possibility of the
construction of the physical system, in which the Mandelbrot set
would arise, was considered. The suggested approach is based on
analysis of a motion of a charged particle in a double-peak
potential with non-linear damping. The particle is driven by
magnetic field, depending on time and on the particle velocity,
and effected by external shot pulses, time intervals between which
also depend on the particle velocity.

In works~\cite{oscillator,oscillator2} and in the present work, we
offer a simpler and universal approach of constructing models
manifesting the Mandelbrot set and other phenomena of CAD, which
may be designed as realistic physical systems. This method allow
us to carry out a physical experiment and present the first
observation of the Mandelbrot set~\cite{Isaeva}. The special
structure of the Fourier spectrum of signal generated by
experimental system at the period-tripling accumulation point is
presented in~\cite{Isaeva2}.

The method developed in~\cite{oscillator} and in the present work
is based on using coupled systems, demonstrating transition to
chaos through period-doublings. It is known that such behavior is
characteristic for a very wide class of nonlinear dissipative
systems of various physical nature.

As shown in~\cite{oscillator}, it easy enough to arrange
realization of the Mandelbrot set for systems with discrete time
and non-autonomous periodically-driven systems (coupled logistic
and H\'{e}non maps and coupled oscillators with quadratic
nonlinearity and harmonic external driving have been
investigated). For this one considers two identical elements
demonstrating the period-doublings with symmetrical coupling,
arising from complexification of the variables and the control
parameter, responsible for the period doubling in the original
system. Domain of generalized partial synchronization (defined by
the term, that the dynamical variables of subsystems do not escape
far from each other) corresponds to the Mandelbrot set.

Such special kind of coupling provides a special symmetry in the
multi-dimensional system, which is necessary for implementation of
the analyticity conditions for the discrete time model map (or the
stroboscopic Poincare map for the continuous system). It is a
simple problem to construct the system with such coupling in
comparison to the system, suggested by Beck.

From our point of view, the problem of realization of phenomena of
CAD for autonomous continuous systems is now opened and seems
interesting. Consideration of this problem is a main goal of the
present paper. In section~2 we reproduce the method of the
obtaining of the coupling function for the logistic maps. In
section~3 we try to implement a similar procedure to a continuous
flow system, namely, to the R\"{o}ssler oscillators.

\section{From complex quadratic map to coupled logistic maps}
Let us start with the notion that one-dimensional complex
quadratic map is equivalent to the  system of two real coupled
quadratic maps with a special type of coupling.

Separation of real and imaginary parts in the complex equation
(\ref{eq1}) yields
\begin{equation}
\label{eq2} z'_{re} \to \lambda _{re} - z_{re}^{2} + z_{im}^{2},
\quad z'_{im} \to \lambda _{im} - 2z_{re} z_{im}.
\end{equation}
Next, we introduce the following designations
\begin{equation}
\label{eq3}
\begin{array}{c}
 x_1=z_{re}+\beta z_{im}, \quad x_2=z_{re}-\beta z_{im}, \\
 \lambda_1=\lambda_{re}+\beta \lambda_{im}, \quad \lambda_2=\lambda_{re}-\beta \lambda_{im}.
 \end{array}
\end{equation}
As a result we obtain a system of two coupled quadratic maps
\begin{equation}
\label{eq4}
\begin{array}{c}
x'_{1} \to \lambda_1-x_{1}^{2}+\varepsilon (x_2-x_1)^2, \\ x'_{2}
\to \lambda_2-x_{2}^{2}+\varepsilon (x_1-x_2)^2, \\
\end{array}
\end{equation}
where $\varepsilon=(1+\beta^2)/4\beta^2$ is the parameter of
coupling. Note a special type of coupling in these equations. It
can be interpreted as an identical simultaneous shift of control
parameters in both partial systems, proportional to the squared
difference of dynamic variables at each step of discrete time.

It is worth nothing that the coefficient
$\varepsilon=(1+\beta^{2})/4\beta^{2}$ for any $\beta$ is larger
than $1/4$. Nevertheless, formally we can investigate
system~(\ref{eq4}) with any $\varepsilon$.

In fig.~2 we present the charts of the parameter plane
$(\lambda_1, \lambda_2)$ for the coupled maps~(\ref{eq4}) at
several values of parameter $\varepsilon$. One can see the usual
Mandelbrot set, rotated by $45^{\circ}$, takes place at
$\varepsilon=0.5$. For $0.25<\varepsilon<+\infty$ we have a
distorted Mandelbrot set on the parameter plane. The cactus leaves
of this Mandelbrot set correspond to existence of periodic motion
of different periods. At $\varepsilon=0.25$, the set on the
parameter plane, for which the point starting from the origin does
not escape to infinity, looks like a set of strips, where the
period doubling cycles occur. At $\varepsilon<0.25$ it transforms
to a rhombus-like structure. At a particular $\varepsilon=0$
(uncoupled logistic maps) it is a square.

The generalization of coupling to the case of
$-\infty<\varepsilon<+\infty$ corresponds to the original
map~(\ref{eq1}), variable and parameter of which are so-called
two-component
numbers~\cite{Lavrentjev,Senn,Fjelstad,Ronveaux,Majernic,Band}.
This is a special algebraic system, which elements are defined as
follows
\begin{equation}\label{twocom}
  z=x+iy, \quad i^{2}=a+ib, \quad \mathrm{where} \quad a,b \in \bf R.
\end{equation}

\begin{figure}
\centerline{\includegraphics[width=0.9\textwidth,keepaspectratio]{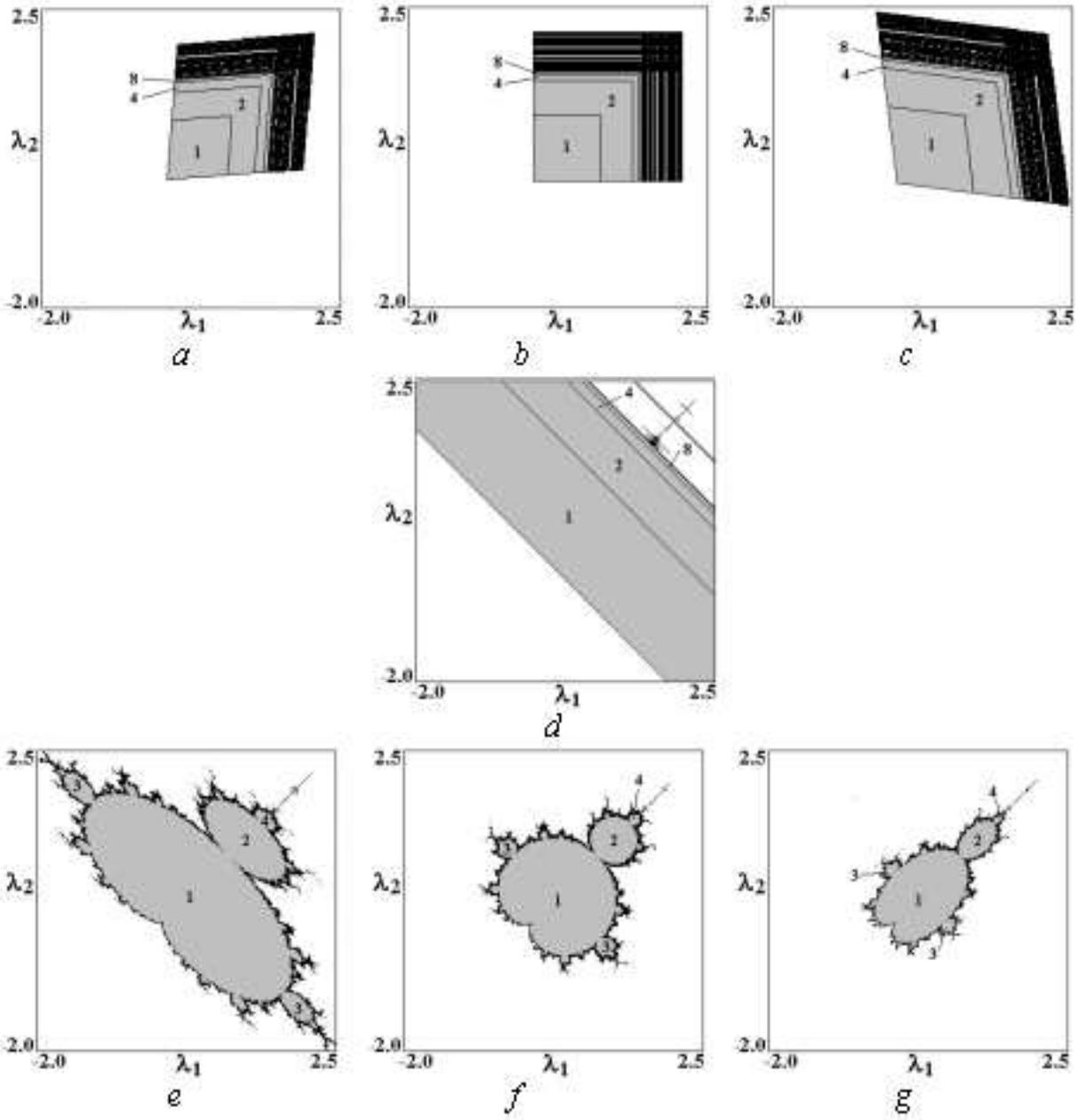}}

\caption{The charts of the parameter plane $(\lambda_1,
\lambda_2)$ for the coupled logistic maps~(\ref{eq4}) with
different values of parameter of coupling: $\varepsilon=-0.1$ (a),
$\varepsilon=0.0$ (b), $\varepsilon=0.1$ (c), $\varepsilon=0.25$
(d), $\varepsilon=0.3$ (e), $\varepsilon=0.5$ (f),
$\varepsilon=1.0$ (g). The figures (a-c) correspond to hyperbolic
numbers, figure (d) -- to parabolic, and figures (e-g) -- to
elliptic numbers.} \label{fig2}

\end{figure}


According to~\cite{Lavrentjev}, there are three special cases:
$i^2=-1$ -- the usual complex numbers, $i^2=+1$ -- the so-called
perplex numbers, $i^2=0$ -- the dual numbers. All other algebraic
number systems are isomorphic to complex, perplex or dual numbers
depending on, whether the value of $(a+b^2)/4b^2$ is positive,
negative or zero, and are known as elliptic, hyperbolic or
parabolic number system, respectively. In terms of parameter
$\varepsilon$ these conditions look as follows:

\begin{description}
\item[1)] the case $\varepsilon>0.25$ corresponds to
elliptic numbers isomorphic to complex numbers, implemented at
$\varepsilon=0.5$;
\item[2)] the case $\varepsilon<0.25$ corresponds to hyperbolic numbers isomorphic to perplex numbers, implemented at
$\varepsilon=0$;
\item[3)] the case $\varepsilon=0.25$ corresponds to parabolic or dual
numbers.
\end{description}

Thus, the existence of three topologically different structures on
the plane of parameters $(\lambda _1,\lambda_2)$, namely, fractal
structure, similar to Mandelbrot set, rhombus-like structure and
system of strips, is explained by existence of three different
algebraic systems of numbers.

\section{Coupled autonomous R\"{o}ssler oscillators}
\begin{figure}
\centerline{\epsfig{file=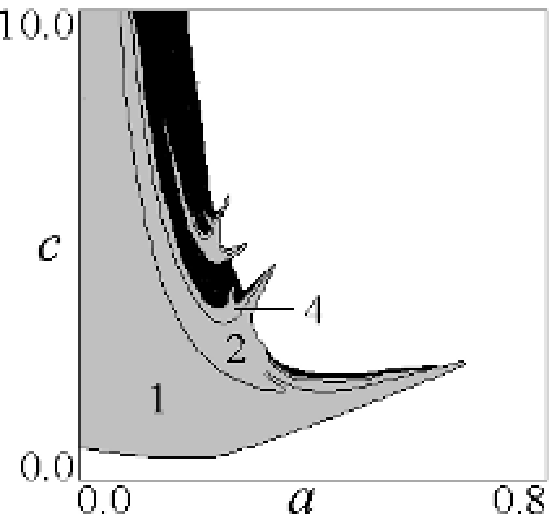,width=0.3\textwidth}\quad
\epsfig{file=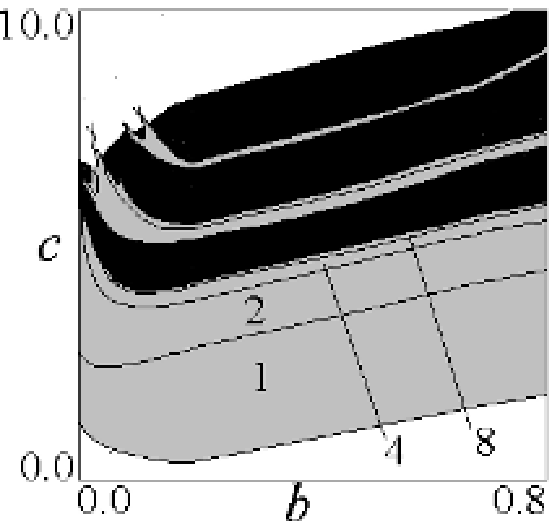,width=0.3\textwidth}\quad
\epsfig{file=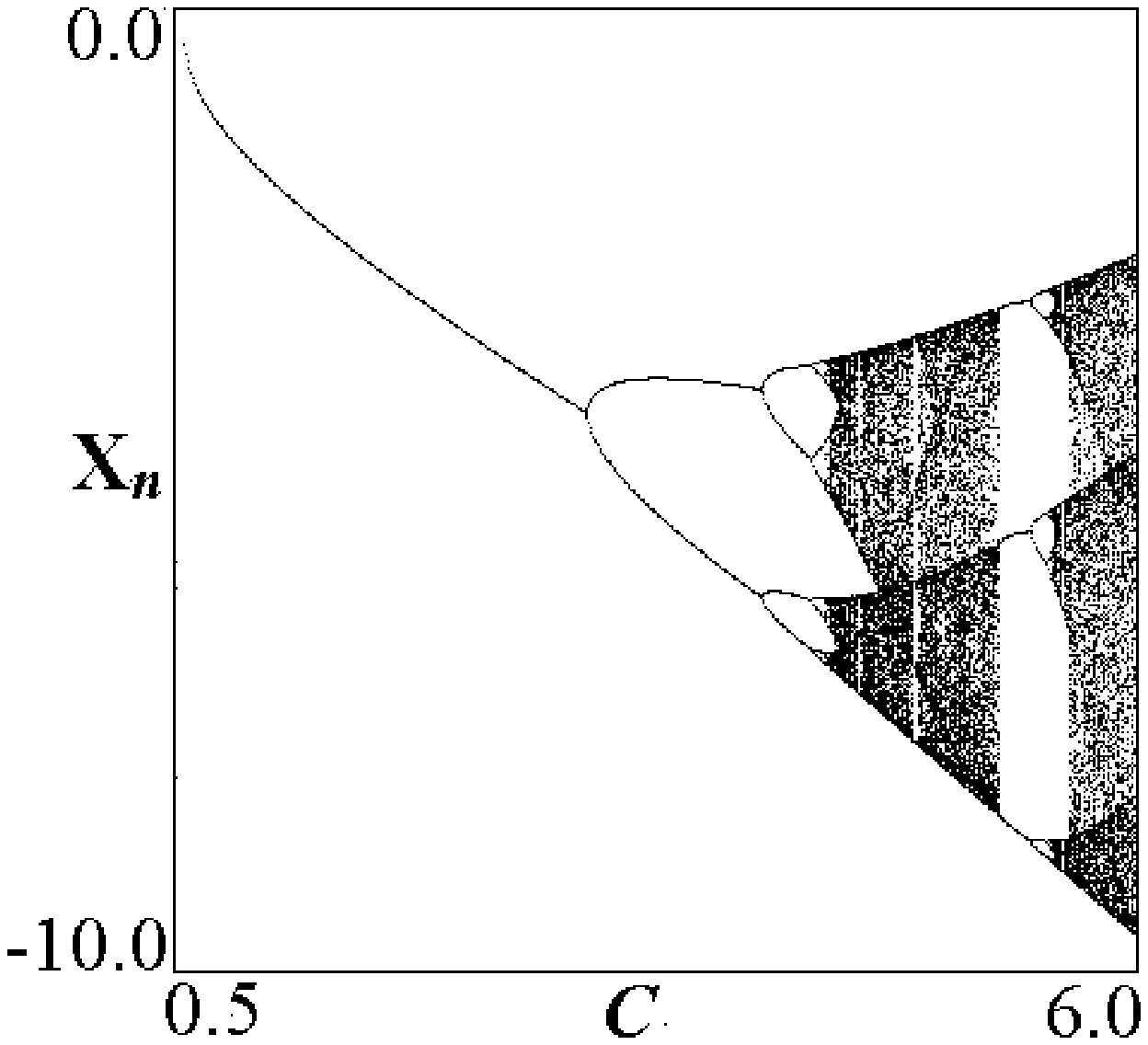,width=0.3\textwidth}}\centerline{(a)\hspace{7cm}(b)}

\caption{Charts of the parameter plane $(c,a)$ with $b=0.2$ (a)
and $(c,b)$ with $a=0.2$ (b) for the R\"{o}ssler oscillator. At
the fig. (c) the bifurcation tree of the R\"{o}ssler oscillator at
the plane $(X_n,c)$ is represented. The value $X_n$ is the
discrete stroboscopic cross-section, corresponded to the values of
dynamic variable $x$ at the Poincare section $y=0$ with $x<0$.
}\label{rom}
\end{figure}

Let us consider autonomous flow system --- R\"{o}ssler oscillator
\begin{equation}
\label{eq12}
\begin{array}{l}
\dot {x}=-(y+z), \\ \dot {y}=x+ay, \\ \dot {z}=b+z(x-c), \\
 \end{array}
\end{equation}
where $x$, $y$, $z$ -- are dynamical variables, $a$, $b$, $c$ --
are parameters~\cite{Rossler}. It is well-known fact, that for the
map at the Poincare cross-section for this system, defined for
example by the plane $y=0$, the transition to chaos through
period-doubling bifurcations is possible. At figure~\ref{rom} the
charts of the parameter $(c,a)$~(a) and $(c,b)$~(b) planes are
shown. Cascade of period-doubling bifurcations is visible, for
example, with changing of parameter $c$ with fixed $a$ and $b$.
Let us take for example the values of parameters $a=0.2$, $b=0.2$.
The bifurcation tree for this values of parameters are represented
at Fig.~\ref{rom}~(c).

Let us construct the system of coupled R\"{o}ssler oscillators. To
obtain the function of coupling we use the procedure described in
previous section. The complexification of R\"{o}ssler system (in
such way, that dynamical variables $x$, $y$, $z$ and parameter,
responsible for the period doublings $c$ are complex, and other
parameters $a$ and $b$ are real) with introduction of the
variables and parameters designations like~(\ref{eq3}) gives the
coupled systems of following form
\begin{equation}
\label{eq13}
\begin{array}{ll}
  \dot{x}_1=-(y_1+z_1), & \dot{x}_2=-(y_2+z_2), \\
  \dot{y}_1=x_1+ay_1, & \dot{y}_2=x_2+ay_2, \\
  \dot{z}_1=b+z_1(x_1-c_1)- &
  \dot{z}_2=b+z_2(x_2-c_2)- \\
  -\varepsilon(z_2-z_1)((x_2-x_1)-(c_2-c_1)), &
  -\varepsilon(z_1-z_2)((x_1-x_2)-(c_1-c_2)).
\end{array}
\end{equation}

But the attempt to obtain the Mandelbrot set at the plane of new
parameters $(c_1,c_2)$ of the system~(\ref{eq13}) has failed. The
investigation of parameter space has not given suitable results --
the domain of generalized synchronization has appeared not to be
similar to Mandelbrot set and is represented by thin band alone
the diagonal line $c_1=c_2$ (this band become more and more thin
with coupling parameter growth). Apparently, the reason of such
behavior lie in the existence of a detuning of the phases of
coupled subsystems. In following subsections we try to avoid the
problem of phase detuning by several special approaches for
construction of the coupled systems.

\subsection{Truncated system of coupled R\"{o}ssler oscillators}

To investigate the synchronization regimes of the coupled
R\"{o}ssler systems often it is more convenient to use a
cylindrical system of coordinates~\cite{Pikovsky}, in which the
set of variables $(x,y,z)$ exchanges by variables ${A,\varphi,z}$,
where $\varphi=\arctan \frac{y}{x}$ -- is a phase,
$A=(x^2+y^2)^{1/2}$ -- is an amplitude. The system~(\ref{eq12}) in
such cylindrical coordinates is represented as follows:
\begin{equation}
\label{eq14}
 \begin{array}{l}
 \dot{A}=aA\sin^2\varphi-z\cos \varphi, \\
 \dot{\varphi}=1+\sin \varphi \cos \varphi+\frac{z}{A}\sin \varphi, \\
 \dot{z}=b-cz+Az\cos \varphi.
 \end{array}
\end{equation}
Poincare cross-section in this case is defined as $\varphi=2\pi
n$, $n=1,2,...$

The coupled R\"{o}ssler oscillators~(\ref{eq13}) rewritten in the
cylindrical coordinates looks as
\begin{equation}
\label{coupRc}
\begin{array}{ll}
  \dot{A}_{1}=aA_1 \sin^{2}\varphi_1-z_1\cos \varphi_1, &
  \dot{A}_{2}=aA_2 \sin^{2}\varphi_2-z_2\cos \varphi_2, \\
  \dot{z}_{1}=b+z_1(A_1 \cos \varphi_1-c_1)- &
  \dot{z}_{2}=b+z_2(A_2 \cos \varphi_2-c_2)- \\
  -\varepsilon(z_2-z_1)((A_2-A_1)\cos \varphi_1-(c_2-c_1)), &
  -\varepsilon(z_1-z_2)((A_1-A_2)\cos \varphi_2-(c_1-c_2)), \\
  \dot{\varphi}_{1}=1+a\sin \varphi_1 \cos \varphi_1+\frac{z_1}{A_1} \sin \varphi_1 &
  \dot{\varphi}_{2}=1+a\sin \varphi_2 \cos \varphi_2+\frac{z_2}{A_2} \sin \varphi_2.
\end{array}
\end{equation}
\begin{figure}
\centerline{\epsfig{file=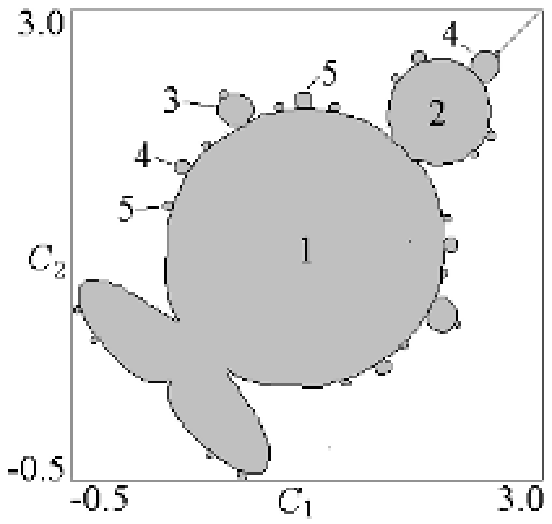,width=0.4\textwidth}}

\caption{Chart of the parameter plane $(c_1, c_2)$ for truncated
system of coupled R\"{o}ssler oscillators (\ref{eq15}) with
$\varepsilon=0.5$, $a=0.2$, $b=0.2$.}\label{fig4}

\centerline{\epsfig{file=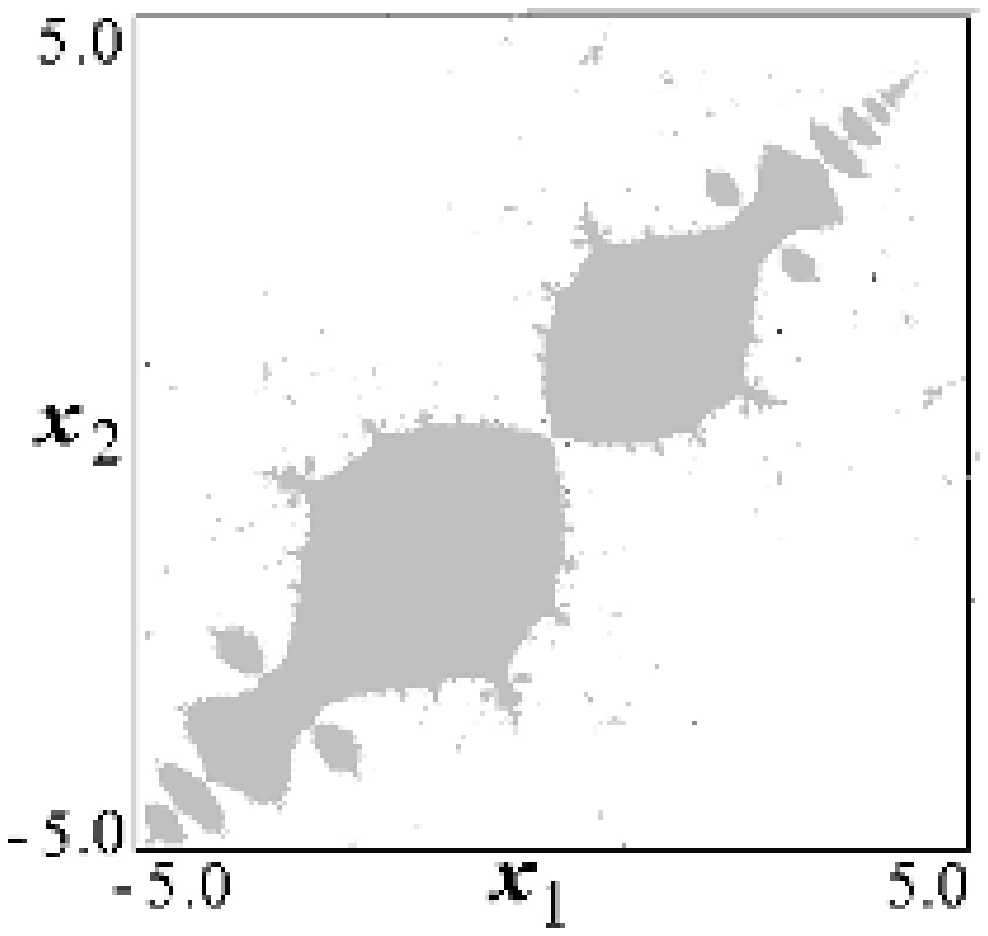,width=0.23\textwidth}\quad
\epsfig{file=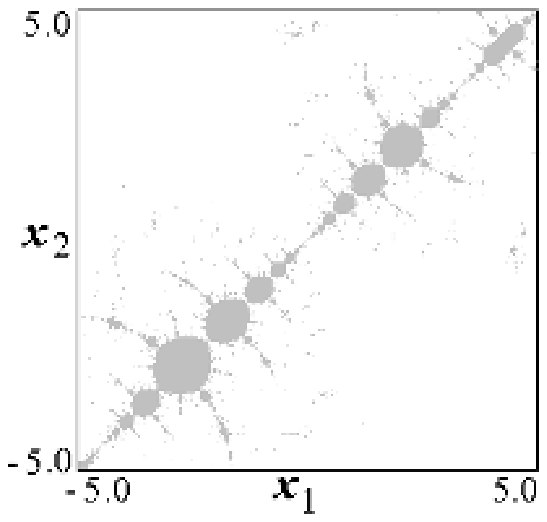,width=0.23\textwidth}\quad
\epsfig{file=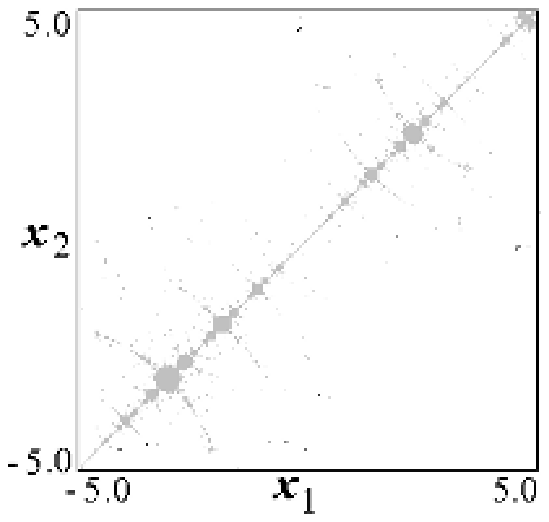,width=0.23\textwidth}\quad
\epsfig{file=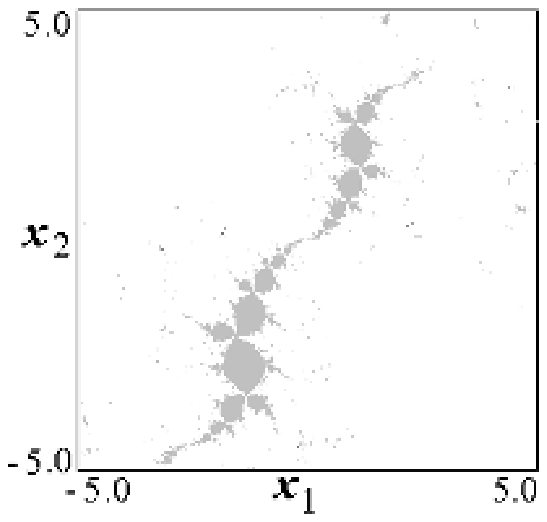,width=0.23\textwidth}}\centerline{(a)\hspace{4cm}(b)\hspace{4cm}(c)\hspace{4cm}(d)}

\caption{Basins of attraction at the phase plane $(x_1=A_1 \cos
\varphi , x_2=A_2 \cos \varphi)$ for truncated system of coupled
R\"{o}ssler oscillators (\ref{eq15}) with $\varepsilon=0.5$,
$a=0.2$, $b=0.2$ for the values of parameters $c_1$ and $c_2$
corresponding to existence of an attracting fixed point at
$c_1=c_2=1.0$ (a), cycle of period 2 at $c_1=c_2=2.25$ (b), cycle
of period 4 at $c_1=c_2=2.58$ (c) and cycle of period 3 at
$c_1=0.75$, $c_2=2.25$ (d) in Poincare cross section.}
\label{fig5}

\centerline{\epsfig{file=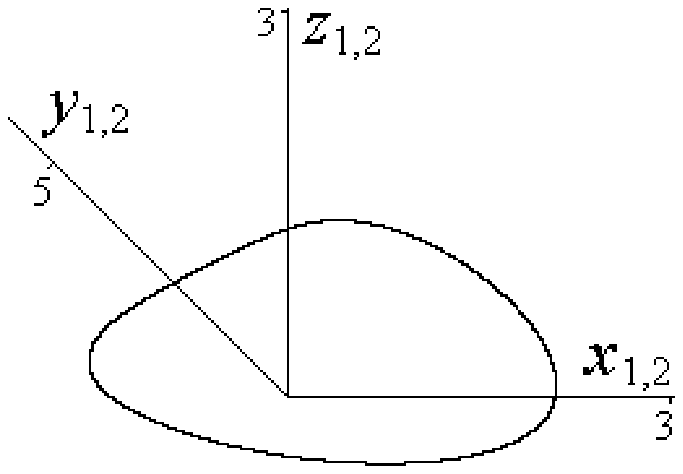,width=0.23\textwidth}\qquad
\epsfig{file=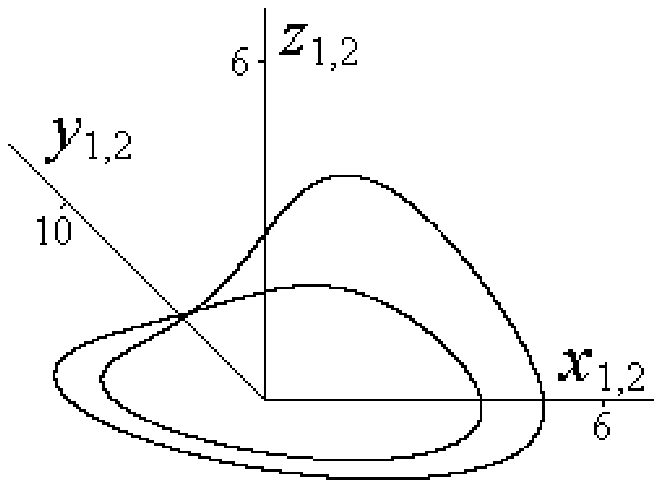,width=0.23\textwidth}\qquad
\epsfig{file=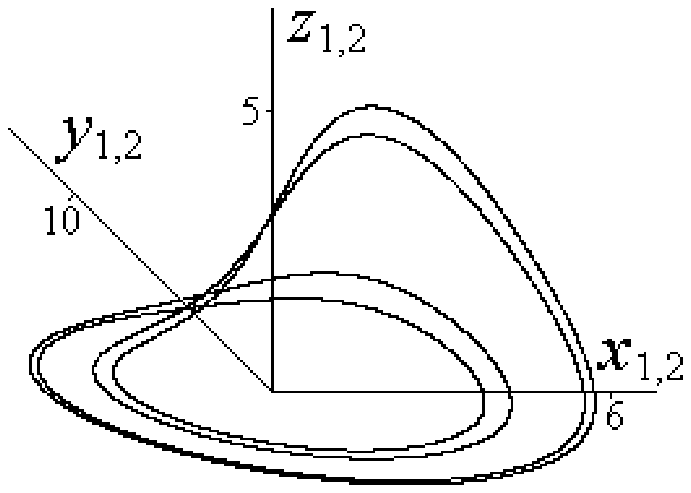,width=0.23\textwidth}}\vspace{-0.3cm}\centerline{(a)\hspace{5cm}(b)\hspace{5cm}(c)}

\centerline{\epsfig{file=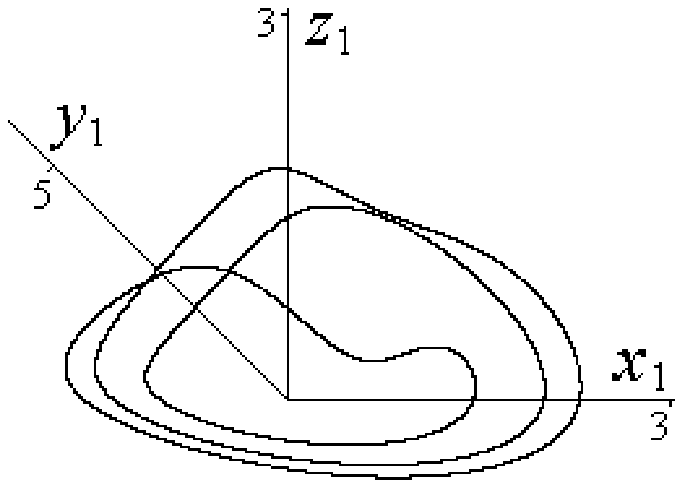,width=0.23\textwidth}\qquad
\epsfig{file=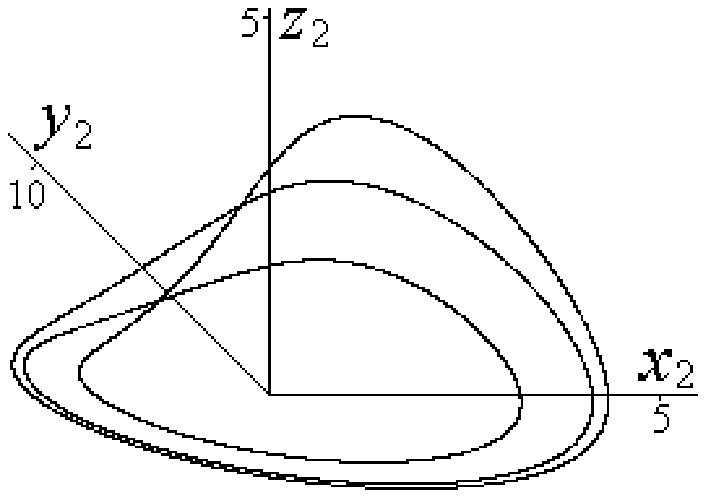,width=0.23\textwidth}}\vspace{-0.3cm}\centerline{(d)\hspace{5cm}(e)}

\caption{Phase portraits for the coupled R\"{o}ssler oscillators
(\ref{eq16}) with the same values of parameters as for
fig.~\ref{fig7}. The attractors of period 1 (a), 2 (b), 3 (c,d)
and 2 (e, f) are represented.} \label{fig55}
\end{figure}
Let us simplify this system, neglecting the second and third terms
at the equations for phases (Such approach does not connect with
some kind of physical interpretation and is considered exclusively
for the showing up the relation between phase synchronization and
possibility of Mandelbrot set realization). In this case, we have
only one variable of phase depending on time by the linear law
$\dot{\varphi}=1$. Then the initial system converts to the form,
similar to nonlinear system with external harmonic driving. Thus,
the simplified system looks as
\begin{equation}
\label{eq15}
\begin{array}{c}
\begin{array}{ll}
  \dot{A}_{1}=aA_1 \sin^{2}\varphi-z_{1} \cos \varphi, &
  \dot{A}_{2}=aA_2 \sin^{2}\varphi-z_{2} \cos \varphi, \\
  \dot{z}_{1}=b+z_1 (A_1 \cos \varphi-c_1)- &
  \dot{z}_{2}=b+z_2 (A_2 \cos \varphi-c_2)- \\
  -\varepsilon (z_2-z_1)((A_2-A_1)\cos \varphi-(c_2-c_1)), &
  -\varepsilon (z_1-z_2)((A_1-A_2)\cos \varphi-(c_1-c_2)),
\end{array} \\
   \dot{\varphi}=1.
\end{array}
\end{equation}
Let us term it as truncated system of coupled R\"{o}ssler
oscillators. It is known from the work \cite{oscillator}, that for
such systems the phenomena of CAD can be realized. Really, the
numerical simulations of~(\ref{eq15}) show, that at a plane of
parameters $(c_1,c_2)$ (fig.~4) one can see the structure similar
to Mandelbrot set, and at the plane $(x_1,x_2)$ (fig.~5) --
structures, similar to Julia sets. The present fact confirms the
guess of necessity of the phase synchronization of coupled
subsystems. At figure~6 the projections of several periodic
attractors (including period-three attractor) in the phase space
are shown.

\subsection{Unidirectionally coupled R\"{o}ssler systems}
Now, let us consider the system of coupled R\"{o}ssler oscillators
with the mutual equation for the phase
\begin{equation}
\label{eq16}
\begin{array}{c}
\begin{array}{ll}
  \dot{A}_{1}=aA_1 \sin^{2}\varphi-z_{1} \cos \varphi, &
  \dot{A}_{2}=aA_2 \sin^{2}\varphi-z_{2} \cos \varphi, \\
  \dot{z}_{1}=b+z_1 (A_1 \cos \varphi-c_1)- &
  \dot{z}_{2}=b+z_2 (A_2 \cos \varphi-c_2)- \\
  -\varepsilon (z_2-z_1)((A_2-A_1)\cos \varphi-(c_2-c_1)), &
  -\varepsilon (z_1-z_2)((A_1-A_2)\cos \varphi-(c_1-c_2)),
\end{array} \\
   \dot{\varphi}=1+\sin\varphi\cos\varphi+\frac{z_1}{A_1}\sin\varphi.
\end{array}
\end{equation}
The similar systems consisting of two coupled subsystems and the
equations, mutual for these subsystems were explored in the works
of Pecora and Caroll~\cite{Pecora1,Pecora2,Caroll}. In these works
the N-dimensional dynamical system $\dot{u}=f(u)$ is subdivided on
two subsystems $\{u\}=\{v,w\}$, one of which is duplicated as
follows $\dot{v}=g(v,w)$, $\dot{w}=h(v,w)$,
${\dot{w}}'=h(v,{w}')$. Then, it is possible to view $\{{w}'\}$ as
response slave system drove by a signal from master system
$\{v,w\}$. The major interest represents the study of
synchronization regimes of this system with a driving
signal~\cite{Pecora1,Caroll}.

The numerical experiment show, that the domain of generalized
synchronization for system~(\ref{eq17}) has some features,
appropriate for CAD. For example, one can see the leave of period
$3$ at the parameter plane (fig.~6). The leaves of higher periods
are small enough and are not visible on the represented chart.

At fig.~7 the basins of attraction at a cross-section plane of a
phase space of system $x_{1}=A_{1}\cos \varphi$, $x_{2}=A_{2} \cos
\varphi$ are represented. Non-symmetrical structure of the domain
of generalized synchronization at fig.~6 and the basins of
attraction at fig.~7, can be explained by unidirectional nature of
coupling between driving and response subsystems.

From a point of view of the experimental physical applications,
dynamics of the system~(\ref{eq17}) requires further study.
\begin{figure}
\centerline{\epsfig{file=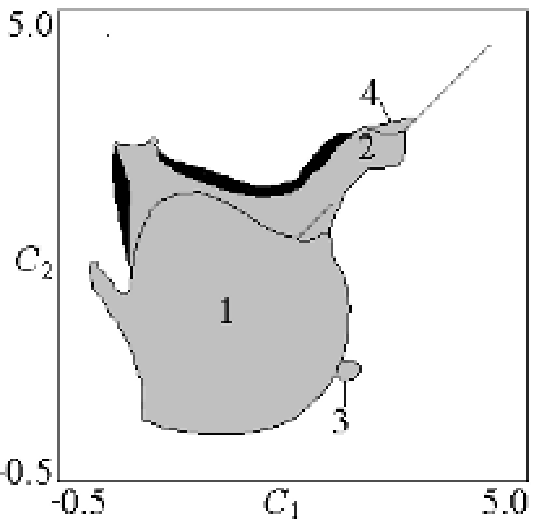,width=0.4\textwidth}}

\caption{Chart of the parameter plane $(c_1, c_2)$ for coupled
R\"{o}ssler oscillators (\ref{eq16}) with $\varepsilon=0.5$,
$a=0.2$, $b=0.2$.} \label{fig6}

\centerline{\epsfig{file=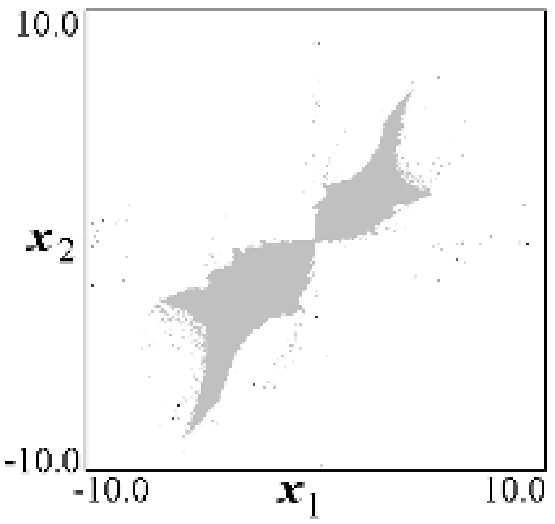,width=0.23\textwidth}\quad
\epsfig{file=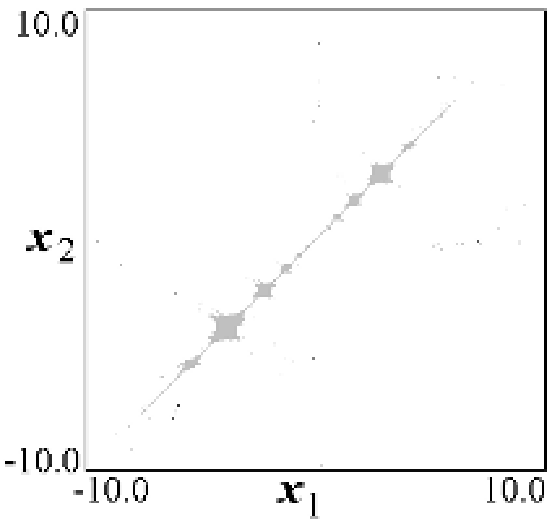,width=0.23\textwidth}\quad
\epsfig{file=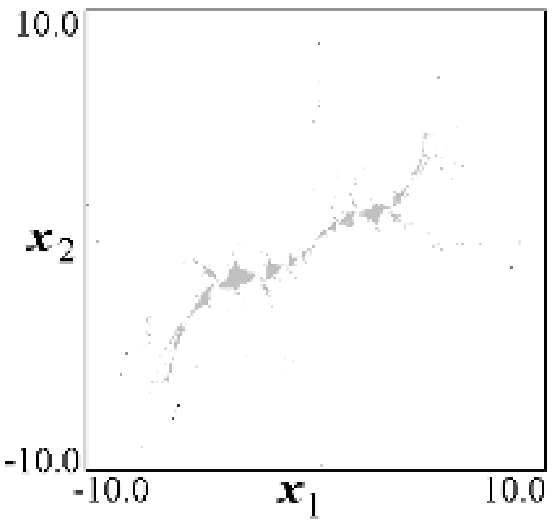,width=0.23\textwidth}\quad
\epsfig{file=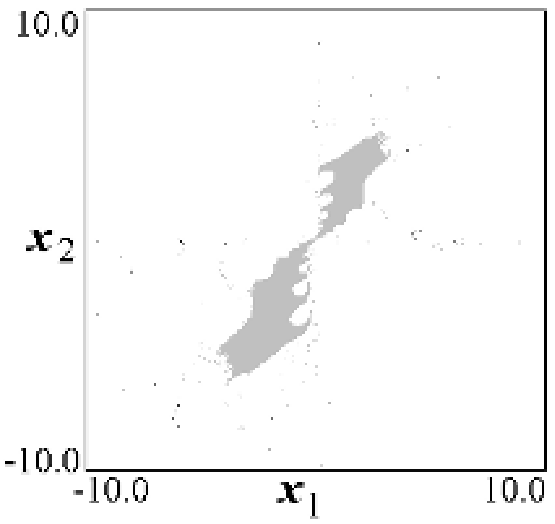,width=0.23\textwidth}}
\centerline{(a)\hspace{4cm}(b)\hspace{4cm}(c)\hspace{4cm}(d)}

\caption{Basins of attraction at the phase plane ($x_1=A_1 \cos
\varphi$, $x_2=A_2 \cos \varphi$) for the system (\ref{eq16}) with
$\varepsilon=0.5$, $a=0.2$, $b=0.2$. The values of parameters
$c_1$ and $c_2$ corresponds to existence of an attracting fixed
point with $c_1=c_2=1.5$ (a), cycle of period 2 with
$c_1=c_2=3.42$ (b), cycle of period 3 with $c_1=2.92$, $c_2=0.83$
(c) and cycle of period 2 at symmetric point $c_1=0.83$,
$c_2=2.92$ (d).} \label{fig7}

\centerline{\epsfig{file=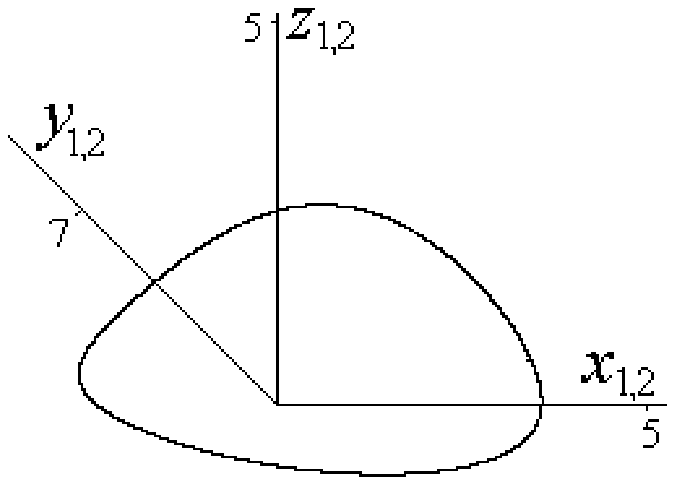,width=0.23\textwidth}\qquad
\epsfig{file=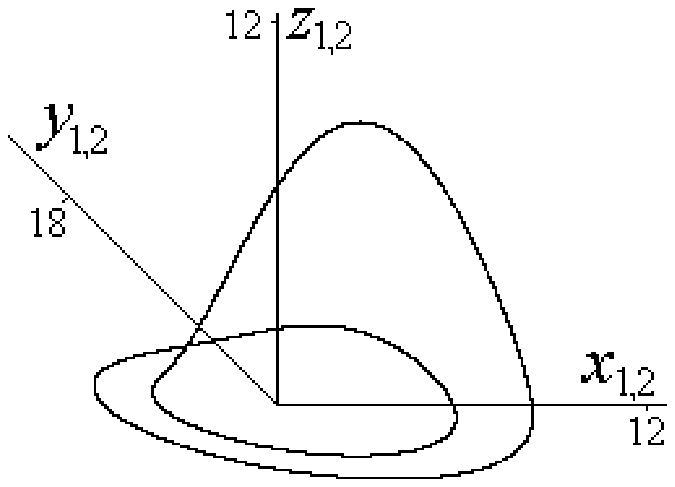,width=0.23\textwidth}}\vspace{-0.3cm}\centerline{(a)\hspace{5cm}(b)}

\centerline{\epsfig{file=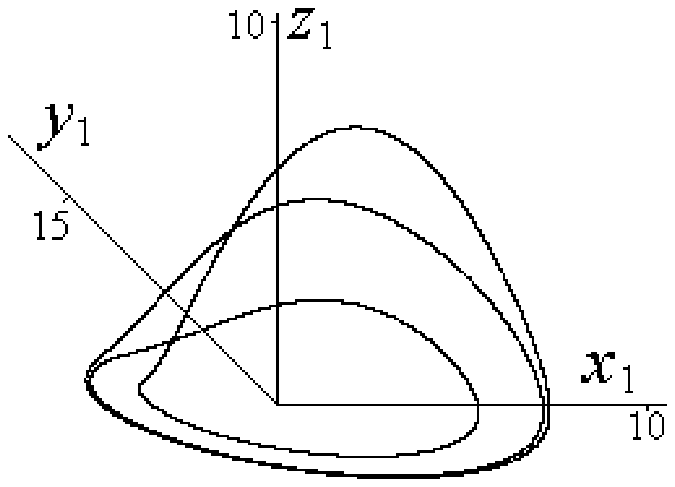,width=0.23\textwidth}\quad
\epsfig{file=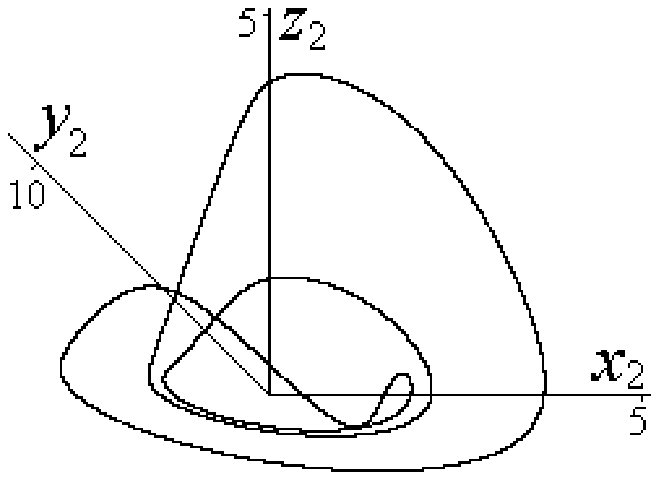,width=0.23\textwidth}\quad
\epsfig{file=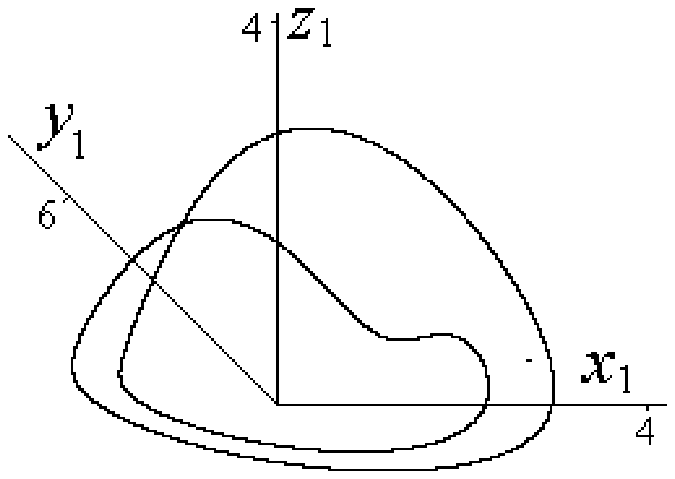,width=0.23\textwidth}\quad
\epsfig{file=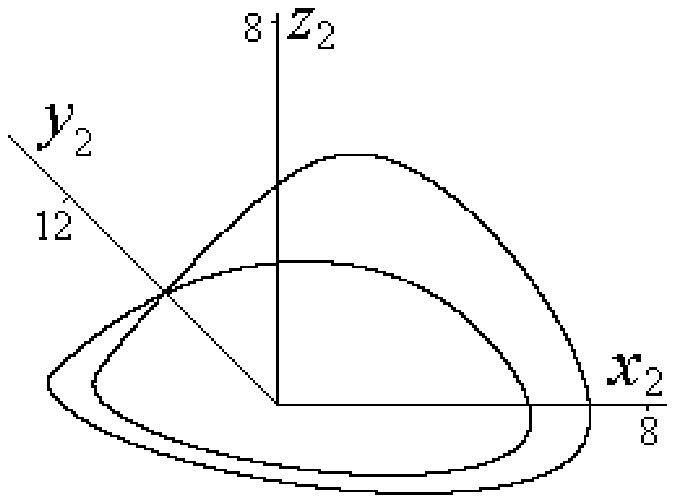,width=0.23\textwidth}}\vspace{-0.3cm}\centerline{(c)\hspace{4.0cm}(d)\hspace{4.0cm}(e)\hspace{4.0cm}(f)}

\caption{Phase portraits for the coupled R\"{o}ssler oscillators
(\ref{eq16}) with the same values of parameters as for
fig.~\ref{fig7}. The attractors of period 1 (a), 2 (b), 3 (c,d)
and 2 (e, f) are represented.} \label{fig110}
\end{figure}

\newpage

\subsection{Coupled R\"{o}ssler oscillators with additional coupling, which synchronize phases}
The next our step is to consider two different equations for
phases $\varphi_{1}$ and $\varphi_{2}$ for each of the partial
systems. To solve the problem of phase detuning we enter the
additional coupling with parameter $\delta$ to the equations for
phases, which thus, provide phase synchronization of R\"{o}ssler
subsystems and does not influence amplitudes
\begin{equation}
\label{eq17}
\begin{array}{ll}
  \dot{A}_{1}=aA_1 \sin^{2}\varphi_1-z_1\cos \varphi_1, &
  \dot{A}_{2}=aA_2 \sin^{2}\varphi_2-z_2\cos \varphi_2, \\
  \dot{z}_{1}=b+z_1(A_1 \cos \varphi_1-c_1)- &
  \dot{z}_{2}=b+z_2(A_2 \cos \varphi_2-c_2)- \\
  -\varepsilon(z_2-z_1)((A_2-A_1)\cos \varphi_1-(c_2-c_1)), &
  -\varepsilon(z_1-z_2)((A_1-A_2)\cos \varphi_2-(c_1-c_2)), \\
  \dot{\varphi}_{1}=1+a\sin \varphi_1 \cos \varphi_1+\frac{z_1}{A_1} \sin \varphi_1 &
  \dot{\varphi}_{2}=1+a\sin \varphi_2 \cos \varphi_2+\frac{z_2}{A_2} \sin \varphi_2 \\
  +\delta \sin(\varphi_2-\varphi_1), &
  +\delta \sin(\varphi_1-\varphi_2).
\end{array}
\end{equation}

In usual rectangular coordinate system ($x=A\cos \varphi$,
$y=A\sin \varphi$, $z$) the equations (\ref{eq17}) looks as
following
\begin{equation}
\label{eq18}
\begin{array}{cc}
  \dot{x}_{1}=-(y_1+z_1)-\delta y_{1}
\frac{x_1y_2-x_2y_1}{{(x_{1}^{2}+y_{1}^{2})^{1/2}(x_{2}^{2}+y_{2}^{2})^{1/2}}},
& \dot{x}_{2}=-(y_2+z_2)-\delta y_{2}
\frac{x_2y_1-x_1y_2}{{(x_{2}^{2}+y_{2}^{2})^{1/2}(x_{1}^{2}+y_{1}^{2})^{1/2}}},
\\
  \dot{y}_{1}=x_1+ay_1+\delta
x_1\frac{x_1y_2-x_2y_1}{(x_1^2+y_1^2)^{1/2}(x_2^2+y_2^2)^{1/2}}, &
  \dot{y}_{2}=x_2+ay_2+\delta
x_2\frac{x_2y_1-x_1y_2}{(x_2^2+y_2^2)^{1/2}(x_1^2+y_1^2)^{1/2}},
\\
  \dot{z}_{1}=b+z_1(x_1-c_1)- & \dot{z}_{2}=b+z_2(x_2-c_2)- \\
  \varepsilon(z_2-z_1)((x_2-x_1)-(c_2-c_1)), &
  \varepsilon(z_1-z_2)((x_1-x_1)-(c_1-c_2)).
\end{array}
\end{equation}

The numerical simulations show, that at a plane $(c_1,c_2)$ for
system (\ref{eq18}) the arising of fractal set, similar to
Mandelbrot set (fig.~8) is possible. Nevertheless, it is easy to
see the considerable distortion of its configuration. For
explanation of this fact we may assume, that the entering of
additional coupling breaks the conditions of complex analyticity
for the resulting stroboscopic Poincare map.

At fig.~9 the basins of attractions similar to Julia sets are
represented. At fig.~10 the attractors of period 1 (a), 2 (b), 4
(c) and 6 (d, e) are shown. Let us note, that with identical
values of parameters $c_1$ and $c_2$ (see fig. a-c) the values of
variables in partial systems are coincided, that corresponds to
the full synchronization. In a case $c_1 \ne c_2$ (fig. d, e),
when the phenomena of CAD can be realized, the values of dynamical
variables of partial systems do not coincide, that corresponds to
realization of generalized partial synchronization. Thus, the
phenomena of CAD such as period-tripling bifurcations can be
implemented, when the point $(c_1,c_2)$ belongs to the region of
generalized partial synchronization.
\begin{figure}
\centerline{\epsfig{file=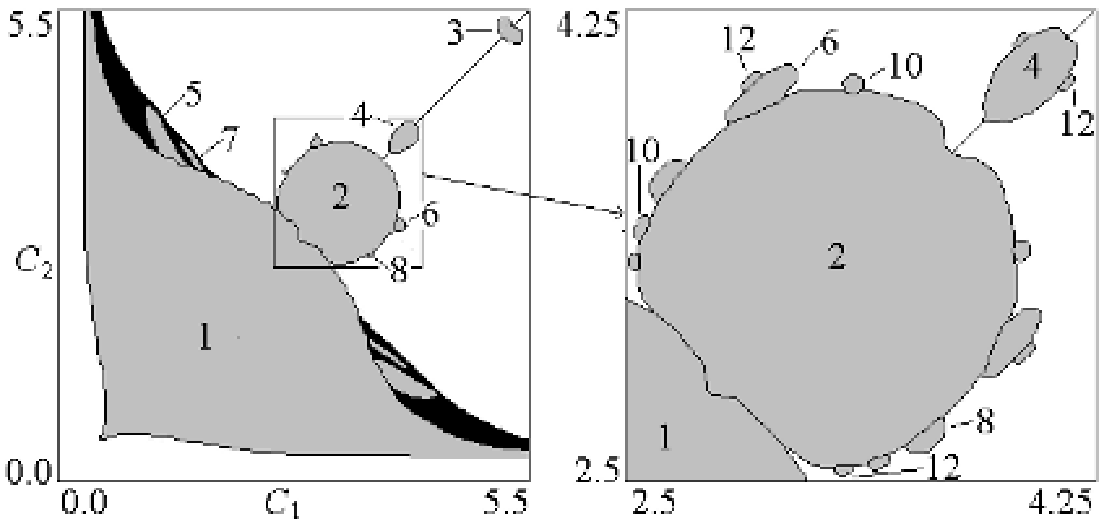,width=0.7\textwidth}}

\caption{Chart of the parameter plane $(c_1,c_2)$ and its enlarged
fragment demonstrating self-similarity for coupled R\"{o}ssler
oscillators (\ref{eq18}) with $\varepsilon=0.5$, $a=0.2$, $b=0.2$,
$\delta=1.0$.} \label{fig8}

\centerline{\epsfig{file=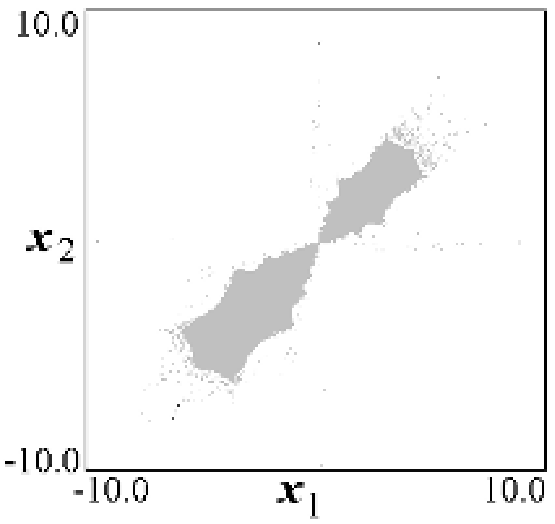,width=0.23\textwidth}\quad
\epsfig{file=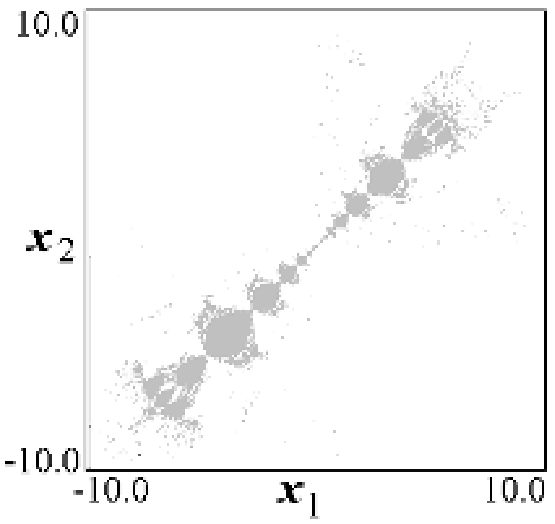,width=0.23\textwidth}\quad
\epsfig{file=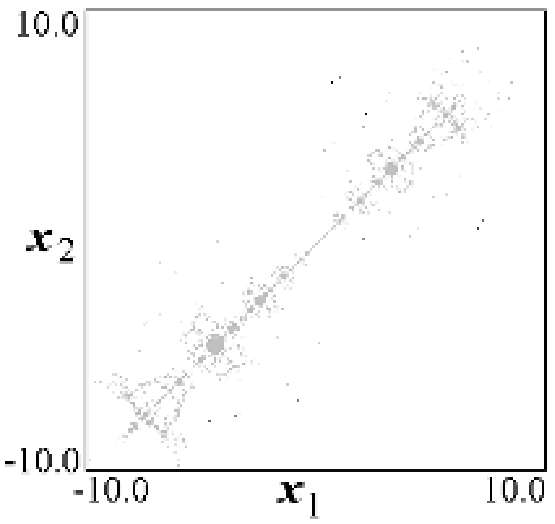,width=0.23\textwidth}\quad
\epsfig{file=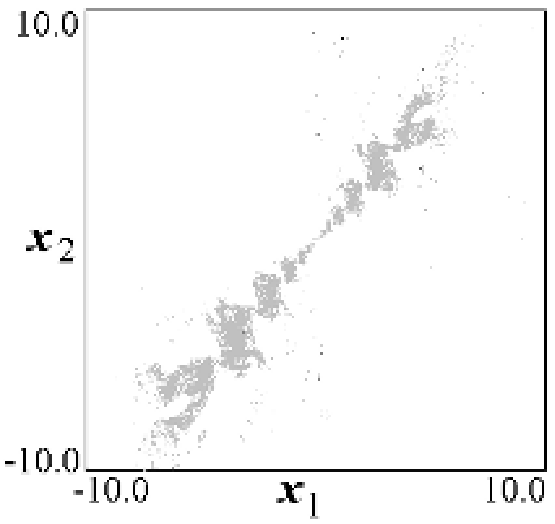,width=0.23\textwidth}}
\centerline{(a)\hspace{4cm}(b)\hspace{4cm}(c)\hspace{4cm}(d)}

\caption{Basins of attraction at the phase plane $(x_1,x_2)$ for
the coupled R\"{o}ssler oscillators (\ref{eq18}) with
$\varepsilon=0.5$, $a=0.2$, $b=0.2$, $\delta=1.0$ for different
values of parameters $c_1$ and $c_2$ corresponding to existence of
an attracting fixed point at $c_1=c_2=2.0$ (a), cycle of period 2
at $c_1=c_2=3.5$ (b), cycle of period 4 at $c_1=c_2=4.0$ (c) and
cycle of period 6 at $c_1=2.98$, $c_2=3.9$ (d).} \label{fig9}

\vspace{0.7cm}

\centerline{\epsfig{file=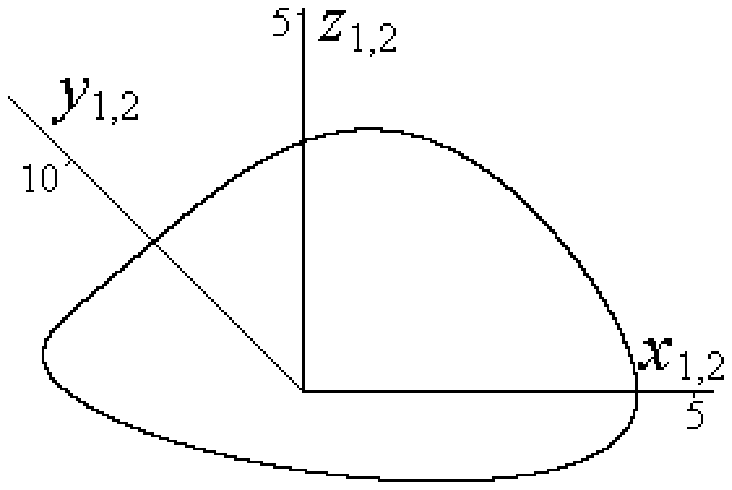,width=0.23\textwidth}\qquad
\epsfig{file=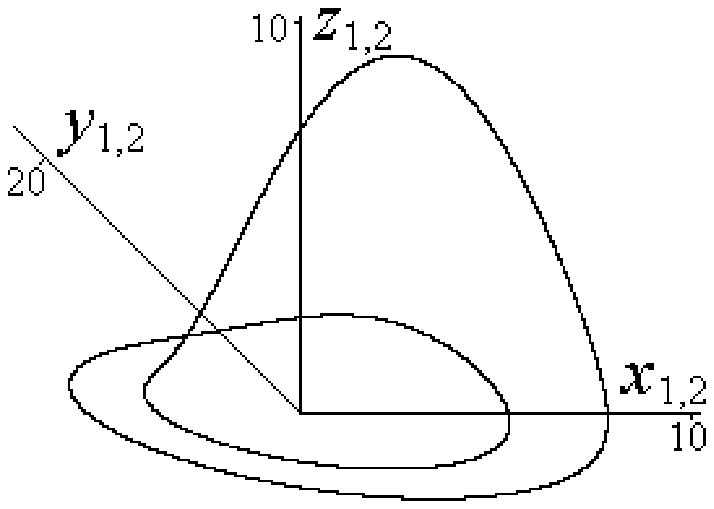,width=0.23\textwidth}\qquad
\epsfig{file=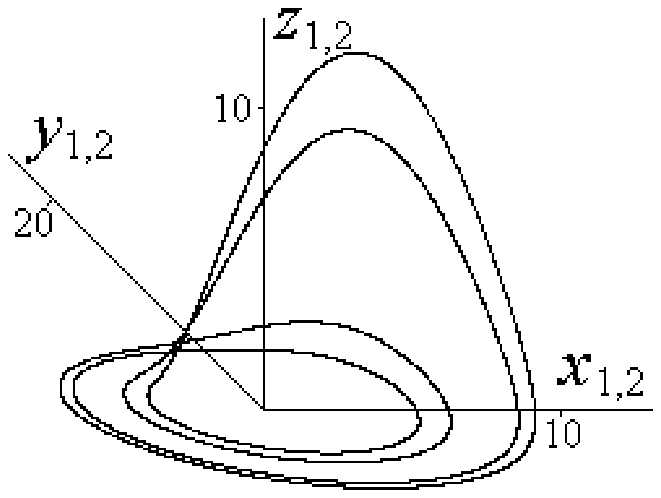,width=0.23\textwidth}}\centerline{(a)\hspace{5cm}(b)\hspace{5cm}(c)}

\centerline{\epsfig{file=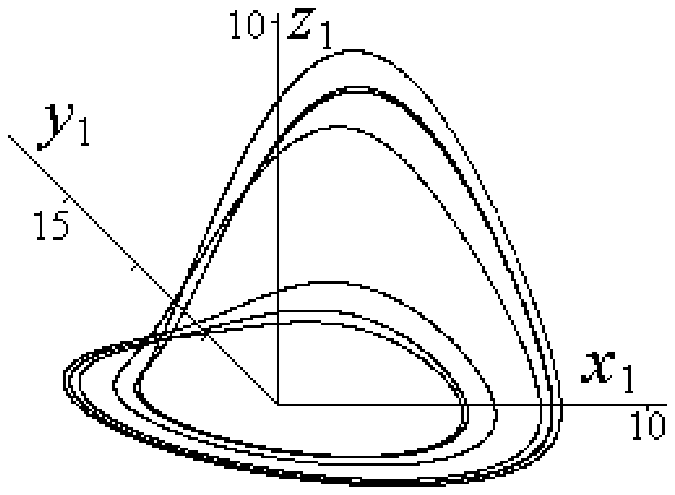,width=0.23\textwidth}\qquad
\epsfig{file=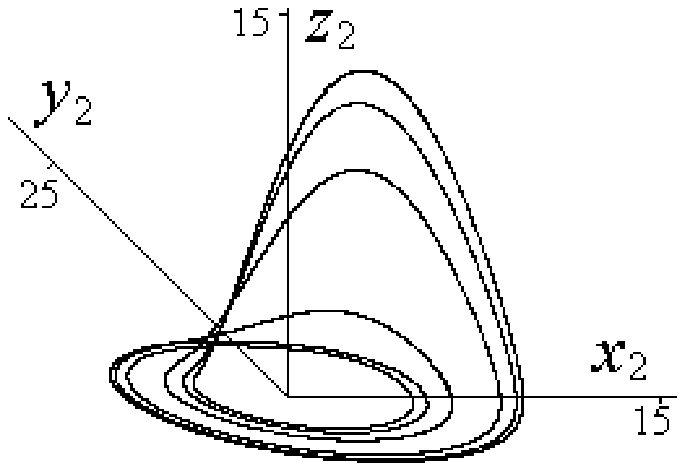,width=0.23\textwidth}}\centerline{(d)\hspace{5cm}(e)}

\caption{Phase portraits for the coupled R\"{o}ssler oscillators
(\ref{eq18}) with the same values of parameters as for fig. 13.
The attractors of period 1 (a), 2 (b), 4 (c) and 6 (d, e) are
represented.} \label{fig10}
\end{figure}
\newpage

\subsection{Coupled R\"{o}ssler oscillators with different time variables}
Let us suggest one more method of obtaining of Mandelbrot set for
autonomous system with continuous time, such as the coupled
R\"{o}ssler oscillators, which allow to avoid the problem of phase
detuning and doesn't break analyticity conditions. This method
lies in following: let us complexify R\"{o}ssler system by another
way. Let us consider the variables $A$, $z$ and time $t$ as
complex and the variable $\varphi$ as real numbers. For this
purpose, it is suitable to rewrite the original R\"{o}ssler system
in cylindrical coordinates as
\begin{equation}
\label{eq20}
 \begin{array}{l}
 \frac{dA}{d\varphi}=\frac{aA\sin^2\varphi-z\cos \varphi}{1+\sin \varphi \cos \varphi+\frac{z}{A}\sin \varphi}, \\
 \frac{dz}{d\varphi}=\frac{b-cz+Az\cos \varphi}{1+\sin \varphi \cos
    \varphi+\frac{z}{A}\sin \varphi}, \\
\frac{dt}{d\varphi}=\frac{1}{1+\sin \varphi \cos
\varphi+\frac{z}{A}\sin \varphi}.
 \end{array}
\end{equation}
After complexification and the variables and parameter
designations we obtain the system of coupled reorganized
R\"{o}ssler oscillators
\begin{equation}
\label{eq22}
 \begin{array}{ll}
 \frac{dA_{1}}{d\varphi}=\frac{A_{1}(aA_{1}\sin^2\varphi-z_{1}\cos \varphi)}{A_{1}+A_{1}\sin \varphi \cos \varphi+z_{1}\sin \varphi}+&
 \frac{dA_{2}}{d\varphi}=\frac{A_{2}(aA_{2}\sin^2\varphi-z_{2}\cos \varphi)}{A_{2}+A_{2}\sin \varphi \cos \varphi+z_{2}\sin \varphi}+ \\
+\varepsilon f_{1}(A_{1},A_{2},z_{1},z_{2},c_{1},c_{2}),&
+\varepsilon f_{1}(A_{2},A_{1},z_{2},z_{1},c_{2},c_{1}), \\

 \frac{dz_{1}}{d\varphi}=\frac{A_{1}(b-z_{1}(c_{1}-A_{1}\cos \varphi))}{A_{1}+A_{1}\sin \varphi
 \cos \varphi+z_{1}\sin \varphi}+&
 \frac{dz_{2}}{d\varphi}=\frac{A_{2}(b-z_{2}(c_{2}+A_{2}\cos \varphi))}{A_{2}+A_{2}\sin \varphi
 \cos \varphi+z_{2}\sin \varphi}+\\
 +\varepsilon f_{2}(A_{1},A_{2},z_{1},z_{2},c_{1},c_{2}),&
 +\varepsilon f_{2}(A_{2},A_{1},z_{2},z_{1},c_{2},c_{1}),\\

\frac{dt_{1}}{d\varphi}=\frac{A_{1}}{A_{1}+A_{1}\sin \varphi \cos
\varphi+z_{1}\sin \varphi}+&
\frac{dt_{2}}{d\varphi}=\frac{A_{2}}{A_{2}+A_{2}\sin \varphi \cos
\varphi+z_{2}\sin \varphi}+ \\

+\varepsilon f_{3}(A_{1},A_{2},z_{1},z_{2},c_{1},c_{2}),&
+\varepsilon f_{3}(A_{2},A_{1},z_{2},z_{1},c_{2},c_{1}).

 \end{array}
\end{equation}
where $f_{1}$, $f_{2}$ and $f_{3}$ are the functions of coupling,
which can be expressed as follows
\begin{equation}
\label{eq23}
 \begin{array}{l}
 f_{1}(A_{1,2},A_{2,1},z_{1,2},z_{2,1},c_{1,2},c_{2,1})= \\
 \qquad =-\frac
 {-2u_{1,2}u_{2,1}(A_{1,2}(w_{1,2}-w_{2,1})+A_{2,1}(w_{1,2}+w_{2,1}))+2u_{1,2}^{2}A_{2,1}(w_{1,2}-w_{2,1})+2u_{2,1}^{2}A_{1,2}(w_{1,2}+w_{2,1})}
 {u_{1,2}(u_{1,2}^{2}+u_{2,1}^{2})}, \\

f_{2}(A_{1,2},A_{2,1},z_{1,2},z_{2,1},c_{1,2},c_{2,1})= \\
 \qquad =-b\frac
 {-2u_{1,2}u_{2,1}(A_{1,2}+A_{2,1})+2u_{1,2}^{2}A_{2,1}+2u_{2,1}^{2}A_{1,2}}
 {u_{1,2}(u_{1,2}^{2}+u_{2,1}^{2})}+ \\

\qquad +\frac
 {-u_{1,2}u_{2,1}(A_{1,2}(-z_{1,2}((c_{1,2}-A_{1,2}\cos \varphi)-(c_{2,1}-A_{2,1}\cos \varphi))+z_{2,1}((c_{1,2}-A_{1,2}\cos\varphi)+(c_{2,1}-A_{2,1}\cos\varphi))))}
  {u_{1,2}(u_{1,2}^{2}+u_{2,1}^{2})}+ \\

\qquad +\frac
 {-u_{1,2}u_{2,1}(A_{2,1}(-z_{1,2}((c_{1,2}-A_{1,2}\cos\varphi)+(c_{2,1}-A_{2,1}\cos\varphi))+z_{2,1}((c_{1,2}-A_{1,2}\cos\varphi)-(c_{2,1}-A_{2,1}\cos\varphi))))}
  {u_{1,2}(u_{1,2}^{2}+u_{2,1}^{2})}+ \\

  +\frac{u_{1,2}^{2}A_{2,1}(z_{1,2}((c_{1,2}-A_{1,2}\cos\varphi)-(c_{2,1}-A_{2,1}\cos\varphi))-z_{2,1}((c_{1,2}-A_{1,2}\cos\varphi)+(c_{2,1}-A_{2,1}\cos\varphi)))}{u_{1,2}(u_{1,2}^{2}+u_{2,1}^{2})}+
  \\

  +\frac{u_{1,2}^{2}A_{1,2}(-z_{1,2}((c_{1,2}-A_{1,2}\cos\varphi)-(c_{2,1}-A_{2,1}\cos\varphi))+z_{2,1}((c_{1,2}-A_{1,2}\cos\varphi)+(c_{2,1}-A_{2,1}\cos\varphi)))}{u_{1,2}(u_{1,2}^{2}+u_{2,1}^{2})}+
  \\

+\frac{2u_{2,1}^{2}A_{1,2}z_{1,2}(c_{1,2}-A_{1,2\cos\varphi})}{u_{1,2}(u_{1,2}^{2}+u_{2,1}^{2})},
\\

 f_{3}(A_{1,2},A_{2,1},z_{1,2},z_{2,1},c_{1,2},c_{2,1})= \\
 \qquad =-\frac
 {-2u_{1,2}u_{2,1}(A_{1,2}+A_{2,1})+2u_{1,2}^{2}A_{2,1}+2u_{2,1}^{2}A_{1,2}}
 {u_{1,2}(u_{1,2}^{2}+u_{2,1}^{2})}.

 \end{array}
\end{equation}
where
\begin{equation}
\label{eq24}
 \begin{array}{l}
 u_{1,2}=A_{1,2}+A_{1,2}\sin \varphi \cos \varphi +z_{1,2}\sin
 \varphi,\\
w_{1,2}=aA_{1,2}\sin^{2}\varphi -z_{1,2}\cos \varphi

 \end{array}
\end{equation}

The represented unusual way of complexification appear to be
successful by some reasons. At first, let us remind that we miss
the problem of phase detuning. Thus, the variable $\varphi$
responsible to the Poincare cross-section ($\varphi=2\pi n$,
$n=1,2,...$) remains real (it is not suitable to consider the
variable $y$ as real, i.e. to consider the derivatives $dx/dy$,
$dz/dy$, $dt/dy$, because $y$ is an non-monotonous function of
time). Second, such complexification does not destroy analyticity.

\begin{figure}
\centerline{\epsfig{file=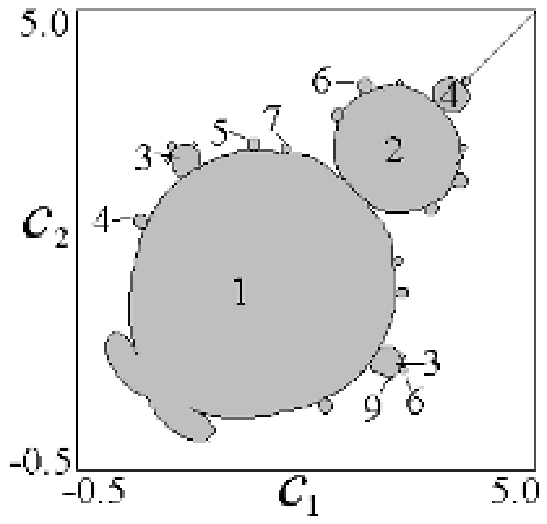,width=0.4\textwidth}}

\caption{The chart of the parameter plane $(c_1,c_2)$ for the
coupled R\"{o}ssler oscillators (\ref{eq22}).} \label{fig11}

\vspace{0.7cm}

\centerline{\epsfig{file=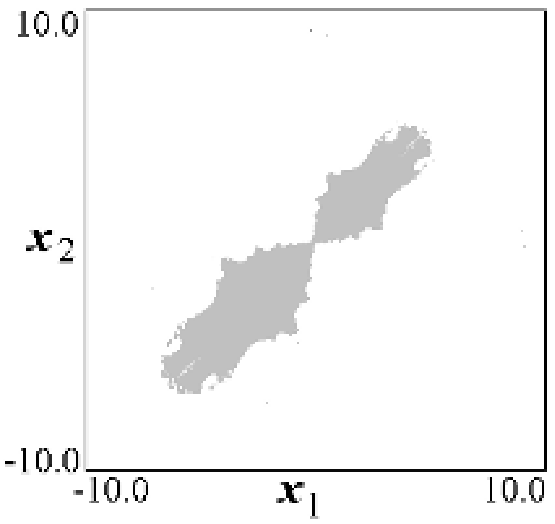,width=0.23\textwidth}\quad
\epsfig{file=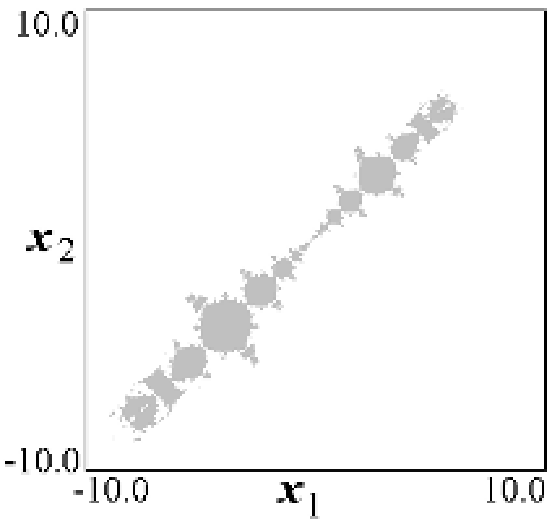,width=0.23\textwidth}\quad
\epsfig{file=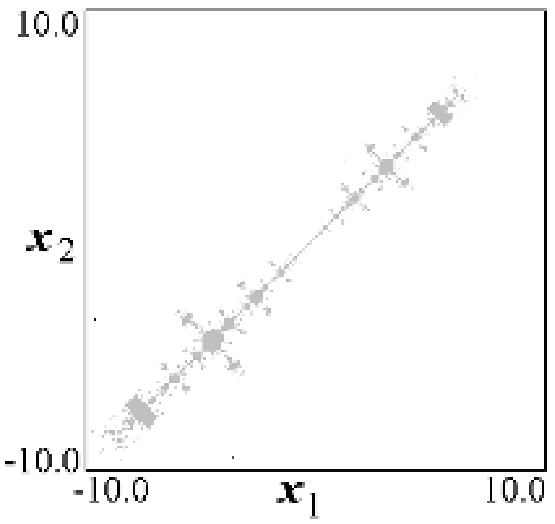,width=0.23\textwidth}\quad
\epsfig{file=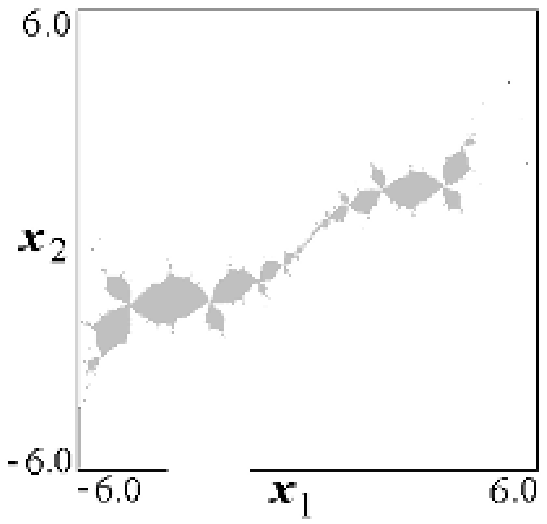,width=0.23\textwidth}\quad}
\centerline{(a)\hspace{4cm}(b)\hspace{4cm}(c)\hspace{4cm}(d)}

\caption{Basins of attraction on the phase planes $(x_1,x_2)$ for
the coupled R\"{o}ssler oscillators (\ref{eq22}). The basins of
attractors of period 1 with $c_1=c_2=1.7$ (a), 2 with
$c_1=c_2=3.35$ (b),4 with $c_1=c_2=4.0$ (c) and 3 with
$c_1=3.222$, $c_2=0.828$ (d) are represented.} \label{fig12}

\vspace{0.7cm}

\centerline{\epsfig{file=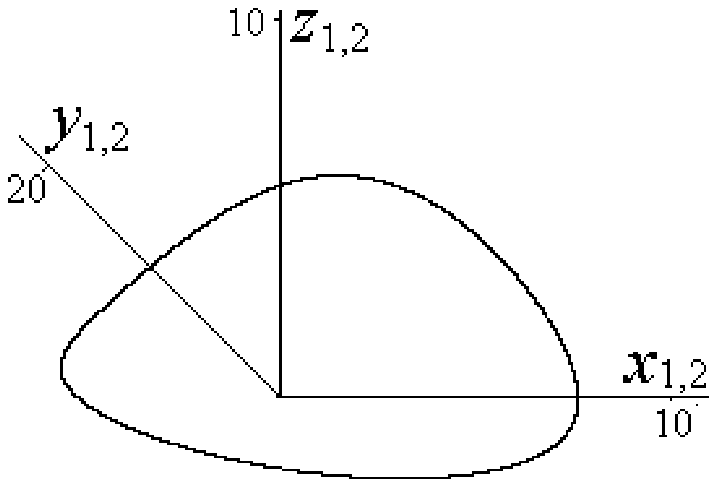,width=0.23\textwidth}\qquad
\epsfig{file=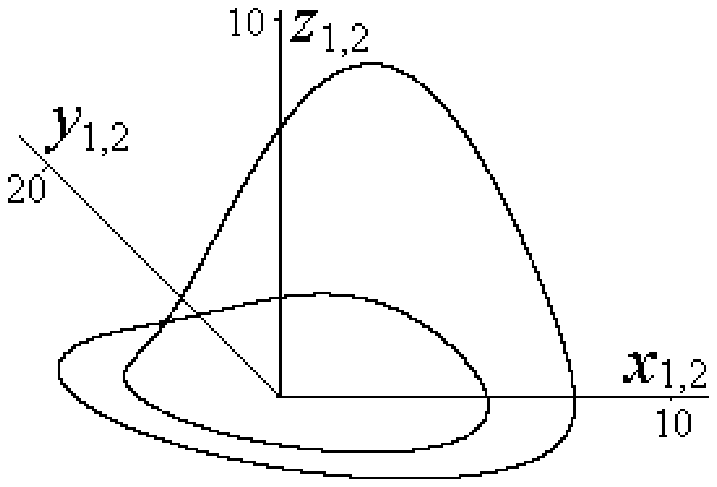,width=0.23\textwidth}\qquad
\epsfig{file=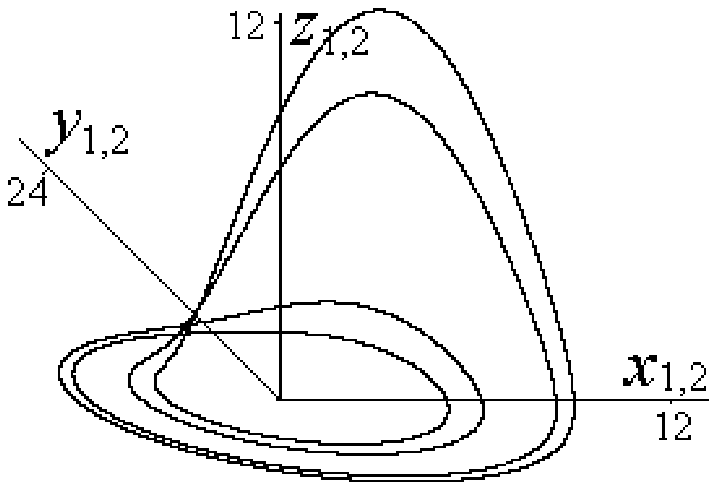,width=0.23\textwidth}}\centerline{(a)\hspace{5cm}(b)\hspace{5cm}(c)}\vspace{0.5cm}

\centerline{\epsfig{file=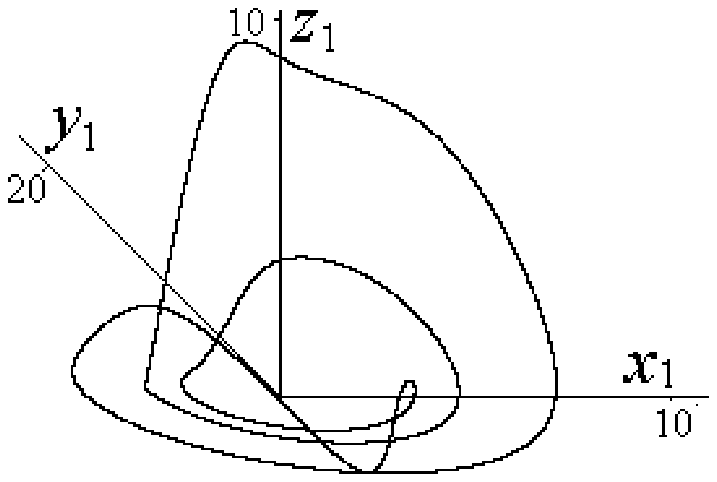,width=0.23\textwidth}\qquad
\epsfig{file=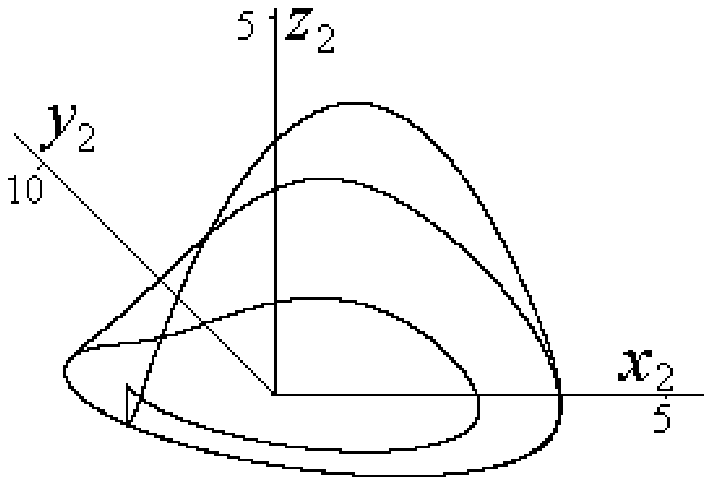,width=0.23\textwidth}}\centerline{(d)\hspace{5cm}(e)}

\caption{Phase portraits for the coupled R\"{o}ssler oscillators
(\ref{eq22}) with the same values of parameters as for fig. 16.
The attractors of period 1 (a), 2 (b), 4 (c) and 3 (d,e) are
represented.} \label{fig13}
\end{figure}

\begin{figure}
\centerline{\epsfig{file=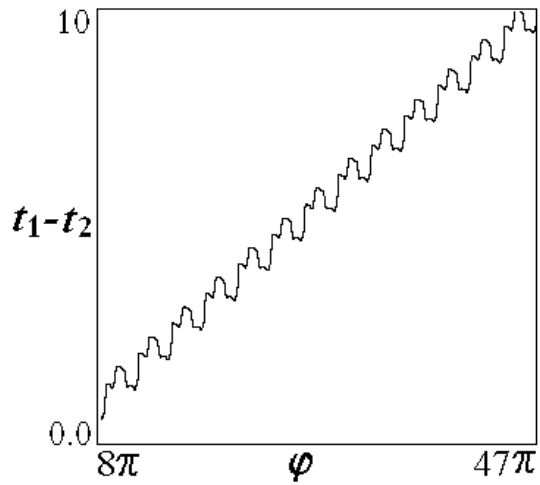,width=0.4\textwidth}}

\caption{The dependence of time detuning on versus the phase with
the values of parameters $c_1=3.222$, $c_2=0.838$ corresponded to
existence of 3-cycle.} \label{fig14}
\end{figure}

The represented at fig.~11 chart of the parameter plane
$(c_1,c_2)$ demonstrates the full agreement with usual Mandelbrot
set. The basins of attraction of cycle of period 1 (fig.~12a), 2
(fig.~12b), 3 (fig.~12c) and 9 (fig.~12d) are also similar with
Julia sets. The next figure~13 shows the attractors for the
subsystems represented in rectangular coordinates. Note, that if
$c_1 \neq c_2$, the variables of time for each subsystem have no
coincided values, i.e. the trajectories of $A_1,z_1,\varphi$ and
$A_2,z_2,\varphi$ run up to the Poincare cross-section
$\varphi=2\pi n$ not simultaneously. Fig.~14 demonstrates the
dependence of time detuning on the variation of phase with the
values of parameters $c_1$, $c_2$ corresponded to cycle of period
3. As it is visible, the value of $t_2-t_1$ oscillates
periodically near the averaged value increasing by linear law.

Also, it is necessary to note, that the variable of time are not
subsisted in the equations for amplitude $A$ and variable $z$, and
doesn't influence its dynamics. Therefore, the equations for $t_1$
and $t_2$ can be rejected. Thus, system~(\ref{eq22}) is also the
truncated system. Besides, unfortunately, it is difficult to
realize the system~(\ref{eq22}) experimentally.

\section{Conclusion}
In present work we have made an attempt to obtain the phenomena of
CAD such as fractal Mandelbrot and Julia sets in the autonomous
continuous systems. As an example the autonomous R\"{o}ssler
oscillator has been considered.

It is shown, that complexified system can be represented as two
coupled real systems. Such representation is useful for some
reasons. At first, it simplify construction of a real physical
model. Secondly, with entering of special coupling to the system
of two identical devices of any nature demonstrating the
period-doubling cascade, one may expect the whole system to
demonstrate phenomena of complex analytic dynamics.

The main result of the work is that the realization of the
Mandelbrot set in parameter space of the autonomous continuous
systems requires provision of additional conditions such as phase
synchronization. The phase synchronization can be obtained by the
truncation of the system of equations, by entering of phase
coupling or by the special type of complexification.

Nevertheless, the problem of the realization of Mandelbrot set in
autonomous flow systems require more careful study. For example,
one can assume, that the codimension of phenomena relevant to CAD
is higher for the flow systems and Mandelbrot set as domain of
generalized synchronization can be observed not on a plane, but on
some surface in parameters space. It is necessary to study in more
detail the regimes of synchronization for the offered system and
the process of influence of a phase detuning on structure of
parameter space.

\section*{Acknowledgements}
This work is supported by RFBR (projects No 03-02-16074 and No 04-02-04011) and CRDF
(REC-006). O.B.I. is grateful to Prof. A. Pikovsky for useful
discussion.


\begin {thebibliography}{99}

\bibitem{Peitgen} H. O. Peitgen, P. H. Richter. The beauty of fractals. Images of
complex dynamical systems. Springer-Verlag. 1986.

\bibitem{Devaney} R. L. Devaney. An Introduction to Chaotic Dynamical Systems. Addison-Wesley
studies in Nonlinearity. 1989.

\bibitem{Golberg} A.I.~Golberg, Y.G.~Sinai, K.M.~Khanin.
Universal properties for sequences of bifurcations of period 3.
Russ.Math.Surv., vol.38, No 1, 1983, P.187-188.

\bibitem{Cvitanovic1} P. Cvitanovi\'{c}, J. Myrheim. Universality for period n-tuplings in
complex mappings // Phys. Lett. A. 1983. V. 94. P. 329.

\bibitem{Cvitanovic2} P. Cvitanovi\'{c}, J. Myrheim. Complex universality // Commun. Math. Phys. 1989.
V. 121. P. 225.

\bibitem{Widom} M. Widom. Renormalization group analysis of quasi-periodicity in analytic
maps // Comm. Math. Phys. 1993. V. 92. P. 121.

\bibitem{Manton} N. S. Manton, M. Nauenberg. Universal scaling bahavior for
iterated maps in the complex plane // Comm. Math. Phys. 1983. V.
89. P. 557.

\bibitem{MacKay} R. S. MacKay, I. C. Percival. Universal small-scale
structure near the boundary of Siegel disks of arbitrary rotation
number. // Physica D. 1987. V. 26. P. 193.

\bibitem{rcd} O.B.~Isaeva, S.P.~Kuznetsov. On scaling properties of
two-dimensional maps near the accumulation point of the
period-tripling cascade. // Regular and Chaotic Dynamics. V. 5,
No. 4, 2000, P. 459-476.

\bibitem{Peinke} J. Peinke, J. Parisi, B. Rohricht, O. E. Rossler // Zeitsch.
Naturforsch. A.\textbf{} 1987. V. 42. P. 263.

\bibitem{Klein} M.Klein. // Zeitsch. Naturforsch. A.\textbf{} 1988. V. 43. P. 819.

\bibitem{Peckham1} B.B. Peckham. Real perturbation of complex
analitic families: Points to regions. // Int.~J. of Bifurcation
and Chaos, V. 8, No.~1, 1998, P. 73-93.

\bibitem{Peckham2} B.B. Peckham. Real continuation from the complex
quadratic family: Fixed-point bifurcation sets. // Int.~J. of
Bifurcation and Chaos, V. 10, No 2, 2000, P.~391-414.

\bibitem{Hu} B. Hu, B. Lin. Yang-Lee zeros, Julia sets, and their singularity spectra //
Phys. Rev. A.\textbf{} 1989. V. 39. P. 4789.

\bibitem{percol} M.~V.~$\mathrm{\acute{E}}$ntin and
G.~M.~$\mathrm{\acute{E}}$ntin, Pis'ma Zh. Eksp. Teor. Fiz.
$\mathbf{64}$, 427 (1996) [JETP Lett. $\mathbf{64}$, 467 (1996)].

\bibitem{Abdusalam} H.A. Abdusalam. Renormalization group method
and Julia sets. // Chaos, Solitons and Fractals, V. 12, 2001, P.
423-428.

\bibitem{npcs} O.B.~Isaeva, S.P.~Kuznetsov. Complex generalization of
approximate renormalization group analysis and Mandelbrot set.
Thermodynamic Analogy. // Nonlinear Phenomena in Complex Systems.
V. 8, No. 2, 2005, P.157-165.

\bibitem{Beck} C. Beck. Physical meaning for Mandelbrot and Julia sets // Physica
D.\textbf{} 1999. V. 125. P. 171.

\bibitem{oscillator} O.B.~Isaeva, S.P. Kuznetzov. On possibility of realization
of the phenomena of complex analytic dynamics in physical systems.
Novel mechanism of the synchronization loss in coupled
period-doubling systems. // Electronic preprint at www.arxiv.org.

\bibitem{oscillator2} O.B.~Isaeva. On possibility of realization of the phenomena of complex analytic
dynamics for the physical systems, built of coupled elements,
which demonstrate period-doublings. // Izv. VUZov PND (Applied
Nonlinear Dynamics). V.~9, No.6, 2001, P. 129-146, (in Russia).

\bibitem{Isaeva} O.B.~Isaeva, S.P.~Kuznetsov, V.I.~Ponomarenko. Mandelbrot set in coupled
logistic maps and in an electronic experiment // Phys. Rev. E.
2001. V. 64. 055201(R).

\bibitem{Isaeva2} O.B.~Isaeva. Universal properties
of the Fourier spectrum of the signal, arising at the
period-tripling accumulation point // Electronic preprint at
www.arXiv.org (submitted to J.~Tech.~Phys.).

\bibitem{Lavrentjev} M.~A.~Lavrentjev and B.~V.~Shabat. Problemy gidrodinamiki
i ikh matematicheskije modeli. (Problems of hydrodynamics and
their mathematical models), Moscow, Nauka, 1977 (in Russian).

\bibitem{Senn} P. Senn. The Mandelbrot set for binary numbers // Am. J. Phys. 1990. V. 58.
P. 1018.

\bibitem{Fjelstad} I. P. Fjelstad. Extending relativity via the perplex numbers // Am. J. Phys.
1986. V. 54. P. 416.

\bibitem{Ronveaux} A. Ronveaux. About `perplex numbers' // Am. J. Phys. 1987. V. 55. P. 392.

\bibitem{Majernic} V. Majernic. The perplex numbers are in fact binary numbers // Am. J. Phys.
1988. V. 56. P. 763.

\bibitem{Band} W. Band. Comments on 'Extending relativity via the perplex numbers' // Am.
J. Phys. 1988. V. 56. P. 469.

\bibitem{Rossler} O. Rossler. An equation for continuos chaos. // Phys. Lett. A. 1976. V. 57.
P.397

\bibitem{Pikovsky} M. G. Rosenblum, A. S. Pikovsky, J. Kurths. From phase to lag
synchronization in coupled chaotic oscillators // Phys. Rev. Lett. 1997. V.
78. P. 4193.

\bibitem{Pecora1} L. M. Pecora, T. L. Caroll. Synchronization in chaotic systems // Phys. Rev.
Lett. 1990. V. 64. P. 821.

\bibitem{Pecora2} L. M. Pecora, T. L. Caroll. Driving systems with chaotic signals // Phys.
Rev. A. 1991. V. 44. P. 2374.

\bibitem{Caroll} T. L. Caroll. Amplitude-independent chaotic synchronisation and
communication // Phys. Rev. E. 1996. V.53. P. 3117.
\end {thebibliography}
\end{document}